\documentclass{aa}

\usepackage{natbib}
\usepackage{graphicx}
\usepackage{amssymb}
\usepackage{txfonts}
\usepackage{xparse,xcolor}
\usepackage[colorlinks=true,linkcolor=blue,citecolor=blue]{hyperref}

\begin{document}

\title{Models of Ultra-Luminous X-ray transient sources}

\author{J.-M. Hameury\inst{1}
    \and
          J.-P. Lasota\inst{2,3}
}

\institute{
     Observatoire Astronomique de Strasbourg, Université de Strasbourg, CNRS UMR 7550, 67000 Strasbourg, France \\
                    \email{jean-marie.hameury@astro.unistra.fr}
\and
     Institut d'Astrophysique de Paris, CNRS et Sorbonne Universit\'e, UMR 7095, 98bis Bd Arago, 75014 Paris, France
\and
     Nicolaus Copernicus Astronomical Center, Polish Academy of Sciences, Bartycka 18, 00-716 Warsaw, Poland
}

   \date{}


  \abstract
  {It is now widely accepted that most ultraluminous X-ray sources (ULXs) are binary systems whose large (above $10^{39}$~erg~s$^{-1}$) apparent luminosities are explained by super-Eddington accretion onto a stellar-mass compact object. Many of the ULXs, especially those containing magnetized neutron stars, are highly variable;
  some exhibit transient behaviour. Large luminosities might imply large accretion discs that could be therefore prone to the thermal--viscous instability known to drive outbursts of dwarf novae and low-mass X-ray binary transient sources.}
   {The aim of this paper is to extend and generalize the X-ray transient disc-instability model to the case of large (outer radius larger than $10^{12}$~cm) accretion discs and apply it to the description of systems with super-Eddington accretion rates at outburst and, in some cases, super-Eddington mass transfer rates.}
   {We have used our disc-instability-model code to calculate the time evolution of the accretion disc and the outburst properties.}
   {We show that, provided that self-irradiation of the accretion disc is efficient even when the accretion rate exceeds the Eddington value, possibly due to scattering back of the X-ray flux emitted by the central parts of the disc on the outer portions of the disc, heating fronts can reach the disc's outer edge generating high accretion rates. We also provide analytical approximations for the observable properties of the outbursts. We have reproduced successfully the observed properties of galactic transients with large discs, such as V404 Cyg, as well as some ULXs such as M51 XT-1. Our model can reproduce the peak luminosity and decay time of ESO 243-39 HLX-1 outbursts if the accretor is a neutron star.}
   {Observational tests of our predicted relations between the outburst duration and decay time with peak luminosity would be most welcome.}


   \maketitle
%

\section{Introduction}

According to the disc instability model \citep[DIM; see][for reviews of the model]{l01,Hameury19}, accretion discs around compact objects are subject to a thermal--viscous instability if the
rate at which matter is brought to their outer edge is less than a critical value strongly increasing with radius. This basic tenet of the DIM has been strongly confirmed by observations of dwarf novae \citep{dohl2018} and transient X-ray sources \citep{c2012}.

Since the critical accretion rate increases with radius, for sufficiently large orbital periods, even potentially very bright systems should have large unstable discs and exhibit some types of outbursts. This could be the case of ultra-luminous X-ray sources \citep[ULXs; see][for a review]{kfr2017}, the majority of which, as now well-established, are apparently super-Eddington luminosity X-ray binaries containing either stellar-mass black-holes or neutron-star accretors \citep[as predicted long time ago by][]{kdwfe2001}. Although it is very difficult to determine the binary parameters of these distant systems (distances up to 20 Mpc) it is clear that a large fraction of them contain giant or supergiant stars\footnote{The companion of a pulsing ULX has been identified in only two cases: a B9 supergiant in NGC 7793 P13 \citep{mpp2014} and a red supergiant in NGC 300 ULX-1 \citep{heida2019}.} and have orbital periods larger than 2 days and thus could have large unstable discs. The recent observation of a very-well sampled transient ULX source in the galaxy M 51 by \citet{bef20} has provided a confirmation of this hypothesis. Since some of the usual, sub-Eddington X-ray transients show long-lasting outbursts (tens of years), this could be also the case of some apparently steady ULXs.  

Despite these potential applications, until recently, modelling the accretion-disc instability has focused on systems with relatively short orbital periods, and disc sizes not exceeding 10$^{11}$cm, with the notable exception of \citet{bhl18} who applied the DIM to two symbiotic stars, Z And and RS Oph that have orbital periods of 759~d and 454~d respectively, implying outer disc radii larger than $10^{12}$cm. \citet{dhl01} also considered discs in LMXBs with radii up to 10$^{12}$~cm, but considered relatively low accretion rates (sub-Eddington).

The reason for the paucity of models with long orbital systems is that modelling large discs is numerically challenging, mainly because of the large ratio between the inner and outer disc radii (up to six orders of magnitude). \citet{lkd15} attempted to describe outbursts of large discs using analytical formulae based on the properties of models for standard X-ray transients (XRTs) but the accuracy of such a method and its application to ULXs should be subject to caution and requires testing through numerical calculations.

The main aim of the present work is to extend and generalize the irradiated-DIM to the case of large (outer radius larger than $10^{12}$~cm) accretion discs and apply it to the description of systems with super-Eddington accretion rates at outburst and, in some cases, super-Eddington mass transfer rates. Such models of ultraluminous X-ray transients can be compared to observations helping both to understand the nature of ULXs and the mechanisms driving accretion in astrophysical discs.

We do not claim, of course, that all (or even most) ULXs are transient systems in which the mass transfer rate from the secondary is sub-Eddington or moderately super-Eddington. Some ULXs, however, \textit{are} observed to be transient, such as for example HLX-1 in ESO 243-49 \citep{fetal2009} or M51 XT-1 \citep{bef20}, and one needs therefore to assess whether the DIM can account for the observed outburst properties, and, if the answer  is yes, under what conditions. On the other hand, many ULXs appear to be permanently in a bright state,  but because of the long time scales involved in large discs, some of them could be caught during outbursts lasting for years or even decades, while other sources, presumably the vast majority if not all, are genuinely permanent\footnote{on timescales of the mass-transfer variations} sources, in which case the DIM sets constraints on the disc size and the mass transfer rate.

As reminded in Sect. \ref{sec:results}, the thermal--viscous stability of discs depends mainly on the mass transfer rate from the secondary, the disc size, the viscosity parameter, and disc irradiation. Our results are therefore, to a very good approximation, independent of the nature of the secondary\footnote{Since we consider here only hydrogen dominated transferred mass, the dependence of the stability criteria on abundances can be here neglected.}. It can be a low-mass or a massive star, it may or may not fill its Roche-lobe; if the mass transfer rate is lower than a critical value depending mainly on the size of the outer disc radius, the system will be transient.

In Sect. \ref{sec:model}, we first briefly summarize the ingredients that enter the DIM as applied to XRTs; we give our results for a grid of models in Sect. \ref{sec:results}, and we show that large accretion rates in outbursts, exceeding the Eddington value can be attained when the mass transfer rate is large enough and the heating front propagates to large distances in the accretion disc, possibly reaching its outer edge. This, however, requires that the irradiation efficiency increases and compensates for the decrease in the accretion efficiency when the Eddington luminosity is exceeded. We then provide in Sect. \ref{sect:analytic} analytical approximations for important properties of the outbursts (peak accretion rate, outburst decay time and duration, etc.) and compare our results with selected observations in Sect. \ref{sect:obs}.

\section{Model} \label{sec:model}

We follow the thermal--viscous evolution of an accretion disc using our code described in \citet{hmd98} and \citet{dhl01}. This code solves the equations for mass and angular momentum conservation, in which the viscosity is parametrized according to the \citet{ss73} prescription, as well as a thermal-balance equation that includes viscous dissipation, heating by tidal torques, and energy advection. Because the disc is geometrically thin, the vertical structure of the disc can be decoupled from its radial variations; the heating and surface cooling terms $Q^+$ and $Q^-$ that enter the thermal equations are known functions of the radius $r$, surface density $\Sigma$, and mid-plane temperature $T_{\rm c}$. We have extended our grid of pre-calculated $Q^+$ and Q$^-$ to radius values of $10^{13}$~cm; at these large distances, the disc temperatures can be very low (below 1000~K) in quiescence, implying that the opacities may have to be extrapolated, and are therefore uncertain. For low temperatures, we use the tables from \citet{a75} that extend down to 700~K; more recent and more accurate table do exist \citep[see e.g.][]{af94}, but it is unclear that the assumptions used for estimating the abundance of grains and molecules are appropriate for accretion discs. We note, however, that, as shown in Sect. \ref{sec:results}, the structure of the distant regions is used only to describe the heating and cooling front propagation when the temperature is larger than typically 10$^4$~K, so that the inaccuracy of the opacities at low temperatures is not really a problem.

The viscosity parameter $\alpha$ is taken to be bimodal, with a value $\alpha_{\rm c}$ on the cold branch, $\alpha_{\rm h}$ on the hot branch and a smooth transition at a temperature $T_{\rm crit}$ that is the average of the turning points $T_{\rm crit}^+$ and $T_{\rm crit}^-$ of the S-curve that describes the equilibrium in the $\Sigma$ -- $T_{\rm c}$ plane. We use the analytical fits given by \citet{dhl01}, and we checked that these fits are also appropriate for much larger radii than considered there.

\subsection{Accretion luminosity}

When the accretion rate $\dot M$ reaches
\begin{equation}
    \dot{M}_{\rm Edd} = 1.3 \times 10^{18} M_1 \; \rm g \, s^{-1},
\label{eq:mdotedd}
\end{equation}
the accretion luminosity corresponds to the Eddington value for an accretion efficiency of 0.1.
$\dot{M}_{\rm Edd}$ is therefore the Eddington accretion rate.
For a 7~M${_\odot}$ black hole, it is equal to $9.1 \times 10^{18}$~g\,s$^{-1}$.

When the accretion rate is super-Eddington, $\dot M(t)$ cannot be considered to represent the X-ray light curve any longer. In calculating light curves of super-Eddington outbursts, we assume that the apparent luminosity is given by \citep{k09}:
\begin{align}
    L_{\rm x} & =  (1+\ln \dot m) \left[1+\frac{\dot m^2}{\tilde b} \right] L_{\rm Edd} \; & {\rm if} \; \dot m \geq 1 & \nonumber \\
              & =  \dot m \, L_{\rm Edd}  \; & {\rm if} \; \dot m < 1,
\label{eq:lx}
\end{align}
In this relation, $\dot m = \dot M / \dot M_{\rm Edd}$, and $b=(1+\dot m^2/\tilde b)^{-1}$ is a beaming term; the larger the beaming parameter $\tilde b$, the larger $b$ and hence the smaller the beaming effect. \citet{k09} found that the beaming factor has the form $b=\tilde b/\dot m^2$ and determined $\tilde b=73$. Here, we substituted the original beaming term $\tilde b /\dot m^2$ which is valid only for $\dot m > \sqrt{\tilde b}$ by $(1+\dot m^2/\tilde b)^{-1}$ in order to get a smooth transition with the case where beaming is negligible.

\citet{k09} determined $\tilde b=73$ using both theoretical arguments and observational constraints; this value is therefore uncertain, but, as shown below, the peak $\dot m$ we obtain in our models is often moderate, typically less than about ten, so that the influence of the beaming term is also moderate. It is also worth noting that, although the comparison with observations requires a relation between $L_{\rm x}$ and $\dot M$, our models are independent of beaming, but depend on the relation between the bolometric luminosity and $\dot M$ in an indirect and, as we see in the next section, hidden way, via the disc irradiation term.

The actual accretion efficiency $\eta_{\rm t}$, defined as the ratio of the true luminosity (without the beaming factor) to $c^2$, is therefore:
\begin{align}
    \eta_{\rm t} & = 0.1 (1+\ln \dot m) / \dot m \; & {\rm if} \; \dot m \geq 1 & \nonumber \\
              \, & =  0.1  \;  &  {\rm if} \; \dot m < 1,
\label{eq:eta}
\end{align}

Equation \ref{eq:lx} also shows that, unless the value of the Eddington luminosity is increased by the presence of a magnetar type magnetic field, beaming is necessary for the luminosity to significantly exceed the Eddington critical value as is the case in the pulsating ultraluminous X-ray sources (PULXs) which contain neutron stars that are very unlikely to be magnetars \citep{kl2019} and have apparent luminosities up to few hundred times larger than the Eddington value \citep[see Table 2 in][]{kl2020}.

\subsection{Outer and inner disc radii}
The outer disc radius is allowed to vary, and is limited by the tidal torques exerted on the outer parts of the disc. Here, we assume, as in \citet{vh08}, that these torques have a steep exponential dependence on radius, so that they are negligibly weak for a radius smaller than the tidal truncation radius $r_{\rm tid}$ and very large beyond that radius, effectively setting $r_{\rm tid}$ as the maximum disc radius. 

We assume that the inner disc is truncated during quiescence because of the formation of a hot, optically thin, and radiatively inefficient flow \citep[see e.g.][]{lny96,nbm97,nm08}. For simplicity, and because the precise dependence on the transition radius between the classical disc and the hot flow is of little importance as far as the time evolution of the disc is concerned --- what matters is the actual  inner disc radius in quiescence just prior to the onset of an outburst --- we assumed that the inner disc is that of an accretion disc truncated by a fictitious magnetic field with effective magnetic moment $\mu$. We considered values of $\mu$ of the order of 10$^{31}$ to 10$^{32}$~G\,cm$^3$ so that the inner disc radius is a few 10$^9$~cm during quiescence. The ``magnetic field'' we use is just a proxy for the unknown but necessary \citep[at least at some phases of the disc's evolution, see e.g.,][]{bz2016,zdm2020} ``evaporation'' mechanism.

In outburst, the inner disc radius should be close to the compact accretor (assuming the neutron star is not strongly magnetized), i.e., $\sim 10^7$\,cm. In the case of very large discs (outer radius $\gtrsim 10^{12}$\, cm) the resulting large radius ratio is numerically too challenging to be implemented. Therefore, as in \citet{dhl01} we limit the inner disc radius to $10^9$\,cm. Since near the outburst peak the inner disc is hot, the viscous timescale at which density perturbations evolve is there short enough for the adjustments of its structure to be instantaneous compared to the outburst evolution timescale as long as the inner radius is low enough. \citet{dhl01} have checked that light-curves produced with inner radii $10^7$ and $10^9$\,cm are practically identical and we expect this to be also true in our case. Truncation at $10^9$\,cm has also the advantage of allowing to avoid problems with the infamous radiation-pressure instability which appears persistently in models \citep[][see, however, \cite{s2016,laketal2019}]{Jsd2013} but seems to be absent in the real Universe since neither steady sources, nor black hole transients decaying from outburst show any characteristic variability when observed in the range of luminosities where this instability is supposed to operate\footnote{With the notable exception of GRS 1915-105, which might be at the upper edge of the presumed instability strip \citep[see e.g.][]{bmkkp97}} \citep[see also][]{gd2004}. In our $10^9$\,cm--truncated models the contribution of radiation pressure never exceed 25\%.

If, during outbursts, the accretion rate exceeds $\dot M_{\rm Edd}$, it is very likely that large outflows are generated at small radii, typically below the spherization radius $r_{\rm sph} = (27/4) \, \dot m \times  GM_1/c^2$ \citep{ss73}. For all the models we calculated, $r_{\rm sph}$ was found to be significantly smaller than 10$^9$cm; anyway it would be impossible to calculate the vertical disc structures for $r < r_{\rm sph}$. However, this effect is taken into account when calculating the luminosity variations using Eq. \ref{eq:lx}.

In three of the models we calculated, the disc temperature in the innermost disc regions exceeded the limits of the pre-calculated grid providing $Q^+$ and $Q^-$; we found it convenient to truncate the disc at a radius of $3 \times 10^9$~cm or $5 \times 10^9$~cm for these two particular models when the accretion rate exceeded $2 \times 10^{18}$~erg\,s$^{-1}$, and we checked on other models that this approximation did not introduce any change in the properties of the light curves.

\subsection{Disc self-irradiation}

Observations show that the outer parts of accretion discs in low-mass X-ray binaries (LMXBs) are irradiated by the central X-ray source \citep{vpmc1994}. This irradiation determines the disc stability criteria \citep{vp1996,c2012} and strongly influences the properties of outburst light curves \citep{dhl01,kr98}. Unfortunately, in spite of several observational and theoretical attempts to discover what are the irradiation mechanism and geometry \citep[it cannot be direct irradiation,][]{dhl01} we still do not know how and by what accretion discs in X-ray binaries are illuminated.

Faced with this situation, we had no choice but to reach for the prescription by \citet{dlh99} that we used twenty years ago \citep{dhl01}. This ansatz has the advantage of being simple, physically motivated and, what is most important, providing a correct stability criterion \citep{c2012}. 

The irradiation formula in question is obtained considering the vertical structure of an accretion disc modified by adding an extra term $\sigma T_{\rm irr}^4$ to the standard boundary condition $F=\sigma T_{\rm s}^4$, where $\sigma$ is the Stefan--Boltzmann constant, $F$ is the vertical heat flux in the disc generated by viscous dissipation, $T_{\rm s}$ is the disc surface temperature, and $T_{\rm irr}$ is the irradiation temperature given by
\begin{equation}
\sigma T_{\rm irr}^4 = \mathcal{C} \frac{\eta_{\rm t} \dot{M} c^2}{4 \pi r^2},
\label{eq:tirr}
\end{equation}
$\mathcal{C}$ contains all the physics of the irradiation process, and $\eta_{\rm t}$ is  defined by Eq. \ref{eq:eta} (except for low accretion rates, see below). \citet{dhl01} found that the light curves of low-mass X-ray transients are reasonably well reproduced, and  \citet{c2012} found that the corresponding stability criterion provides the observed division of sources into steady and outbursting if one uses a constant $\eta_{\rm t} \mathcal{C}$ of the order of $10^{-3}$. We stress out that our definition of $\mathcal{C}$ is the same as in \citet{dhl01}, but differs from the one used in \citet{dlh99} in that it does not include the efficiency $\eta_{\rm t}$. In the following, we use
\begin{equation}
    f_{\rm irr} = \frac{\eta_{\rm t} \mathcal{C}}{5 \times 10^{-4}}
\end{equation}
to quantify the effect of irradiation.

As in \citet{dhl01}, $\eta_{\rm t}$ is reduced at low accretion rates, because for those low rates, the accretion flow below the truncated disc becomes radiatively inefficient. We use here a cut-off term $[1+(\dot{M}/10^{16} \; \rm g \, s^{-1})^{-4}]^{-1}$. 

For large mass accretion rates, we consider two extreme cases: a.) $C$ is constant below the Eddington luminosity, and is reduced by a factor $(1 + \ln \dot m)$ above it, and b.) $\eta_{\rm t} \mathcal{C}$ is constant at all accretion rates. Case a.) corresponds to a constant irradiating flux above the Eddington value, while case b.) implies that the decrease in $\eta_{\rm t}$ is compensated by an equivalent increase in $\mathcal{C}$. We also consider an intermediate case where $C$ is constant (case a1).

The increase in $\mathcal{C}$ could be due to strong thermal outflows driven by X-ray irradiation of the outer accretion disc that would scatter efficiently the X-rays emitted in the vicinity of the accreting compact object, as proposed by \citet{ddt19}. They estimated in a self-consistent manner the efficiency of disc irradiation from a wind model. However, their approach can apply to sub-Eddington rates only and we consider systems with larger orbital separation than theirs. In addition, when the model is applied to an observed system, the conclusion is that scattering in the wind is still not sufficient to produce the observed heating, even in combination with direct illumination \citep{tdmdc2020}. Here, since discs \textit{are observed} to be irradiated, we adopt a simple (even simplistic) ``pragmatic'' approach in order to produce models that can be tested by observations, as has been the case of X-ray transient models tested by \citet{tdl18,tlh18}.
 
\section{Results} \label{sec:results}
In the following, the primary is either a black hole with mass $M_1$=7~M$_\odot$ or a neutron star with $M_1$=1.4~M$_\odot$; the secondary mass is in all but one cases $M_2$=0.4~M$_\odot$. We consider three possible values for the orbital period $P_{\rm orb}$: 155~h, 400~h, and 1200~h, corresponding to an outer disc radius of approximately 10$^{12}$, $2 \times 10^{12}$, and $4 \times 10^{12}$~cm respectively in the black hole case, and to $5 \times 10^{11}$, $10^{12}$, and $2 \times 10^{12}$~cm when the accretor is a neutron star. The value of $M_2=0.4 \, \rm M_{\odot}$ is obviously not suitable for mass transfer rates up to $3 \times 10^{-7} \rm M_{\odot}/yr$ but we use this mass for the models homogeneity. We are anyway interested in considering the impact of the disc radius on the outcome of the model, and not on modelling a specific system with the exception of V404 Cyg and M51 XT-1. When one modifies both the secondary mass and the orbital period in such a way that the outer disc radius remains the same, the only disc parameter that is different is the specific angular momentum of matter incorporated at the outer edge of the disc, which is usually parametrized by the so-called circularization radius $r_{\rm circ}$. As we show below (models 34 and 34a in Sec. \ref{sec:sequ}), the outburst properties are essentially independent of $r_{\rm circ}$. The secondary mass we have chosen corresponds to the properties of V404 Cyg, the binary parameters of M51 X-1 are not known. We finally note that if the disc is fed via a wind from the secondary instead of Roche lobe overflow, the outer disc radius is smaller than what we infer from the orbital separation; again, because of the very weak dependence on $r_{\rm circ}$, our results are valid also in this case provided one considers the actual value of the disc radius.

In the following, the mass transfer rate is taken as a free parameter; in real systems, it will be determined by the orbital separation and the secondary properties in a complex way that we do not address here \citep[see e.g.][for a discussion of the various formation channels of ULXs]{wsl17}. But even in the case where the secondary is more massive than the primary and experiences mass loss on a thermal time scale, the disc can in principle  still be unstable if it is large enough.

\begin{table*}
\caption{Model parameters for black hole accretors}
\begin{tabular}{lllllllllllllll}
\hline \hline
Nr. & $P_{\rm orb}$ & $\dot{M}_2$ & $f_{\rm irr}$ & $\mu_{30}$ & $\alpha_{\rm c}$  & $t_{\rm c}$ & $t_{\rm r}$ & $d$ & $\dot{M}_{\rm max}$ & $r_{\rm tr,12}$ & $\Delta M$ & $\Delta M/M_{\rm d}$ &Pattern  \\
    & (hr) & (g\,s$^{-1}$) & & & & (yr) & (yr) & (yr)& (g s$^{-1}$) &  & (g) \\
\hline
1  &155  &5 10$^{16}$ & 1 &  10 & 0.04 & 11.9 &11.9 &0.92 & 1.42 10$^{18}$ & .362 & 1.74 10$^{25}$ & 0.040 &L \\
2  &155  &5 10$^{17}$ & 1 &  10 & 0.04 & 51.8 &13.0 &2.87 & 1.38 10$^{19}$ & 1.00  & 5.12 10$^{26}$ & 0.479 &Lsmm\\
3  &155  &5 10$^{17}$ & 1 &  10 & 0.02 & 39   &39   &2.71 & 1.60 10$^{19}$ & 1.00  & 5.76 10$^{26}$ & 0.489 &L \\
4  &155  &5 10$^{16}$ & 0           &  10 & 0.04 & 1.71 &1.71 &0.24\tablefootmark{(a)} & 1.58 10$^{18}$ & .147 & 2.28 10$^{24}$ & 0.003 &L\\
5  &155  &5 10$^{17}$ & 0           &  10 & 0.04 & 1.60 &1.60 &0.41\tablefootmark{(a)} & 9.94 10$^{18}$ & .267 & 2.07 10$^{25}$ & 0.006 &L \\
6  &400  &5 10$^{16}$ & 1 &  10 & 0.04 & 12   &12   &0.92 & 1.42 10$^{18}$ & .361 & 1.73 10$^{25}$ & 0.014 &L  \\ 
7  &400  &5 10$^{17}$ & 1 &  10 & 0.04 & 25.9 &13   &2.19 & 1.23 10$^{19}$ & 1.01  & 3.28 10$^{26}$ & 0.078 &Lm\\
8  &400  &5 10$^{16}$ & 1 &  10 & 0.02 & 36   &36   &1.13 & 3.70 10$^{18}$ & .551 & 5.64 10$^{25}$ & 0.024 &L\\
9  &400  &7 10$^{16}$ & 1 &  10 & 0.02 & 36.6 &36.6 &1.25 & 4.65 10$^{18}$ & .617 & 7.87 10$^{25}$ & 0.028 &L \\
10  &400  &1 10$^{17}$ & 1 &  10 & 0.02 & 36.8 &36.8 &1.40 & 5.98 10$^{18}$ & .697 & 1.13 10$^{26}$ & 0.033 &L\\
11  &400  &2 10$^{17}$ & 1 &  10 & 0.02 & 74.6 &37.3 &1.93 & 1.47 10$^{19}$ & 1.06  & 3.68 10$^{26}$ & 0.077 &Lm \\
12 &400  &5 10$^{17}$ & 1 &  10 & 0.02 & 448  &37.4 &4.02 & 4.99 10$^{19}$ & 1.95  & 3.02 10$^{27}$ & 0.445 &Ls5*m2*(sm)s\\
13 &400  &7 10$^{17}$ & 1 &  10 & 0.02 & 226  &37.7 &4.24 & 5.86 10$^{19}$ & 1.96  & 3.49 10$^{27}$ & 0.482 &Ls4*m \\
14\tablefootmark{b} &400  &5 10$^{18}$ & 1 &  10 & 0.02 & 75.6 &37.8 &5.49 & 2.01 10$^{20}$ & 1.99  & 1.11 10$^{28}$ & 0.742 &Ls  \\
15 &1200 &5 10$^{16}$ & 1 &  10 & 0.04 & 11.9 &11.9 &0.92 & 1.41 10$^{18}$ & .36   & 1.80 10$^{25}$ & 0.005 &L \\
16 &1200 &5 10$^{16}$ & 4 &  10 & 0.04 & 21.3 &21.3 &1.96 & 1.15 10$^{18}$ & .516 & 3.07 10$^{25}$ & 0.011 &L \\
17 &1200 &5 10$^{16}$ & 4  & 100 & 0.04 & 29.2 &29.2 &2.04 & 1.52 10$^{18}$ & .587 & 4.26 10$^{25}$ & 0.014 &L  \\
18 &1200 &5 10$^{16}$ & 1 &  10 & 0.02 & 36.1 &36.1 &1.13 & 3.71 10$^{18}$ & .551 & 1.46 10$^{26}$ & 0.019 &L \\
19 &1200 &5 10$^{17}$ & 1 &  10 & 0.04 & 26   &13   &2.18 & 1.24 10$^{19}$ & 1.01  & 3.28 10$^{26}$ & 0.024 &Lm \\
20 &1200 &5 10$^{17}$ & 1 &  10 & 0.02 & 72.9 &36.4 &2.61 & 3.26 10$^{19}$ & 1.55  & 1.09 10$^{27}$ & 0.041 &Ls\\
21 &1200 &2 10$^{18}$ & 1 &  10 & 0.02 & 147  &36.7 &4.27 & 1.40 10$^{20}$ & 3.08  & 7.52 10$^{27}$ & 0.147 &L$\mu$m$\mu$\\
22\tablefootmark{b} &1200 &5 10$^{18}$ & 1 &  10 & 0.02 & 334  &37.1 &8.01 & 3.67 10$^{20}$ & 4.20  & 3.76 10$^{28}$ & 0.574 & L$\mu$s3*(m$\mu$)\\
23 \tablefootmark{c} &1200 &2 10$^{19}$ & 1 &  10 & 0.02 & 77.5 &38.7 &8.74 & 4.68 10$^{20}$ & 4.21  & 4.41 10$^{28}$ & 0.605 & L$\mu$\\
24 &1200 &5 10$^{18}$ & 1\tablefootmark{(d)} &  10 & 0.02 & 67.3 &67.3 &31.8 & 8.04 10$^{19}$ & 1.23  & 5.72 10$^{27}$ & 0.076 &L& \\
24a &1200 &5 10$^{18}$ & 1\tablefootmark{(d)}&  10 & 0.02 & 40.4 & 40.4 & 4.01 & 1.48 10$^{20}$ & 2.03 & 5.85 10$^{27}$ & 0.052 &L&\\
\hline
\end{tabular} \\
\tablefoottext{a}{reflares during decay}
\tablefoottext{b}{inner disc truncated at $3 \times 10^{9}$~cm near outburst maximum}
\tablefoottext{c}{inner disc truncated at $5 \times 10^{9}$~cm near outburst maximum}
\tablefoottext{d}{irradiation limited at Eddington luminosity; see text.}
\end{table*}

\begin{table*}
\caption{Model parameters for neutron star accretors}
\begin{tabular}{lllllllllllllll}
\hline \hline
Nr. & $P_{\rm orb}$ & $\dot{M}_2$ & $f_{\rm irr}$ & $\mu_{30}$ & $\alpha_{\rm c}$  & $t_{\rm c}$ & $t_{\rm r}$ & $d$ & $\dot{M}_{\rm max}$ & $r_{\rm tr,12}$ & $\Delta M$ & $\Delta M/M_{\rm d}$ &Pattern  \\
    & (hr) & (g\,s$^{-1}$) & & & & (yr) & (yr) & (yr)& (g\,s$^{-1}$) &  & (g) \\
\hline
25 & 155 & 5 10$^{16}$ & 1 & 10 & 0.02 & 25.2 & 25.2 & 0.73 & 3.43 10$^{18}$ & .422 & 3.78 10$^{25}$ & 0.425 & L \\
26 & 155 & 5 10$^{17}$ & 1 & 10 & 0.02 & 15.8 & 15.8 & 1.25 & 2.55 10$^{19}$ & .468 & 2.30 10$^{26}$ & 0.813 & L \\
27 & 155 & 2 10$^{18}$ & 1 & 10 & 0.02 & 10.0 & 10.0 & 2.07 & 8.17 10$^{19}$ & .461 & 5.18 10$^{26}$ & 0.891 & L \\
28 & 400 & 5 10$^{16}$ & 1 & 10 & 0.02 & 26.5 & 26.5 & 0.71 & 4.00 10$^{18}$ & .462 & 4.02 10$^{25}$ & 0.089 & L \\
29 & 400 & 7 10$^{16}$ & 1 & 10 & 0.02 & 27.3 & 27.3 & 0.80 & 5.14 10$^{18}$ & .524 & 5.81 10$^{25}$ & 0.112 & L \\
30 & 400 & 1 10$^{17}$ & 1 & 10 & 0.02 & 28.2 & 28.2 & 0.90 & 6.76 10$^{18}$ & .602 & 8.57 10$^{25}$ & 0.147 & L \\
31 & 400 & 2 10$^{17}$ & 1 & 10 & 0.02 & 90.0 & 30.0 & 1.52 & 1.48 10$^{19}$ & .887 & 3.46 10$^{26}$ & 0.513 & Lmm \\
32 & 400 & 5 10$^{17}$ & 1 & 10 & 0.02 & 29.7 & 29.7 & 1.50 & 2.26 10$^{19}$ & .94 & 4.46 10$^{26}$ & 0.580 & L \\
33 & 400 & 7 10$^{17}$ & 1 & 10 & 0.02 & 28.3 & 28.3 & 1.62 & 3.12 10$^{19}$ & .935 & 5.92 10$^{26}$ & 0.644 & L \\
34 & 1200 & 5 10$^{17}$ & 1 & 10 & 0.02 & 33.5 & 33.5 & 1.58 & 2.28 10$^{19}$ & 1.11 & 5.05 10$^{26}$ & 0.101 & L \\
34a\tablefootmark{a} & 1505& 5 10$^{17}$ & 1 & 10 & 0.02 & 33.5 & 33.5 & 1.58 & 2.28 10$^{19}$ & 1.11 & 5.05 10$^{26}$ & 0.102 & L \\
35 & 1200 & 2 10$^{18}$ & 1 & 10 & 0.02 & 41.1 & 41.1 & 2.60 & 6.52 10$^{19}$ & 1.94 & 2.50 10$^{27}$ & 0.501 & L \\
\hline 
\end{tabular} \\
\tablefoottext{a}{$M_2=5$~M$_\odot$}
\end{table*}

\begin{figure}
\includegraphics[width=\columnwidth]{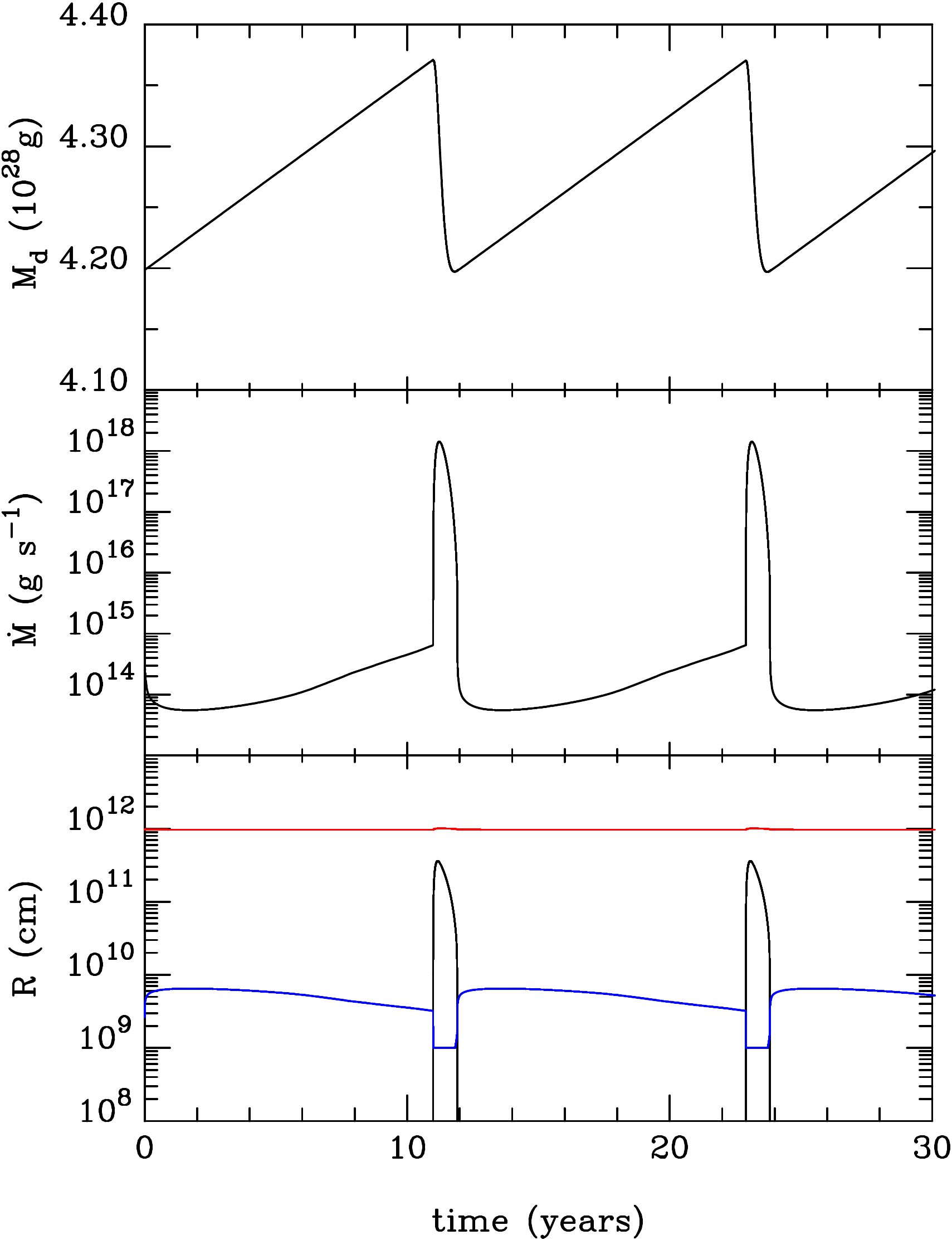}
\caption{Time evolution for model 1, which corresponds to parameters appropriate for V404 Cyg. The upper panel represents the disc mass, the intermediate panel the mass accretion rate, and the bottom panel the outer disc radius (red curve), the inner disc radius (blue curve), and the transition radius $r_{\rm tr}$ (black curve). For such sub-Eddington outbursts mass-accretion variations represent the X-ray light-curve.}
\label{fig:model1}
\end{figure}

\begin{figure}
\includegraphics[width=\columnwidth]{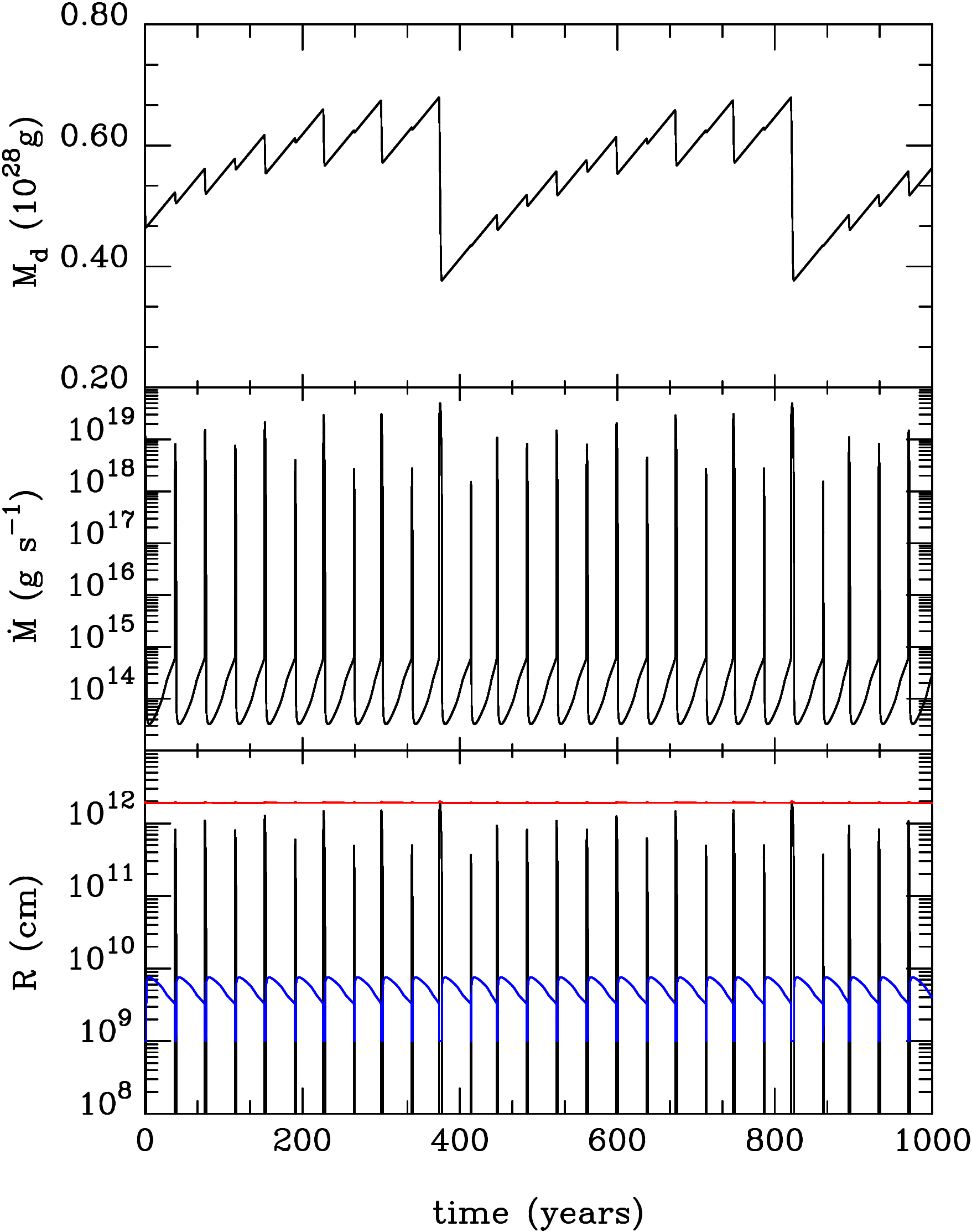}
\caption{Same as Fig. \ref{fig:model1} for model 12.}
\label{fig:model12}
\end{figure}\begin{figure}
\includegraphics[width=\columnwidth]{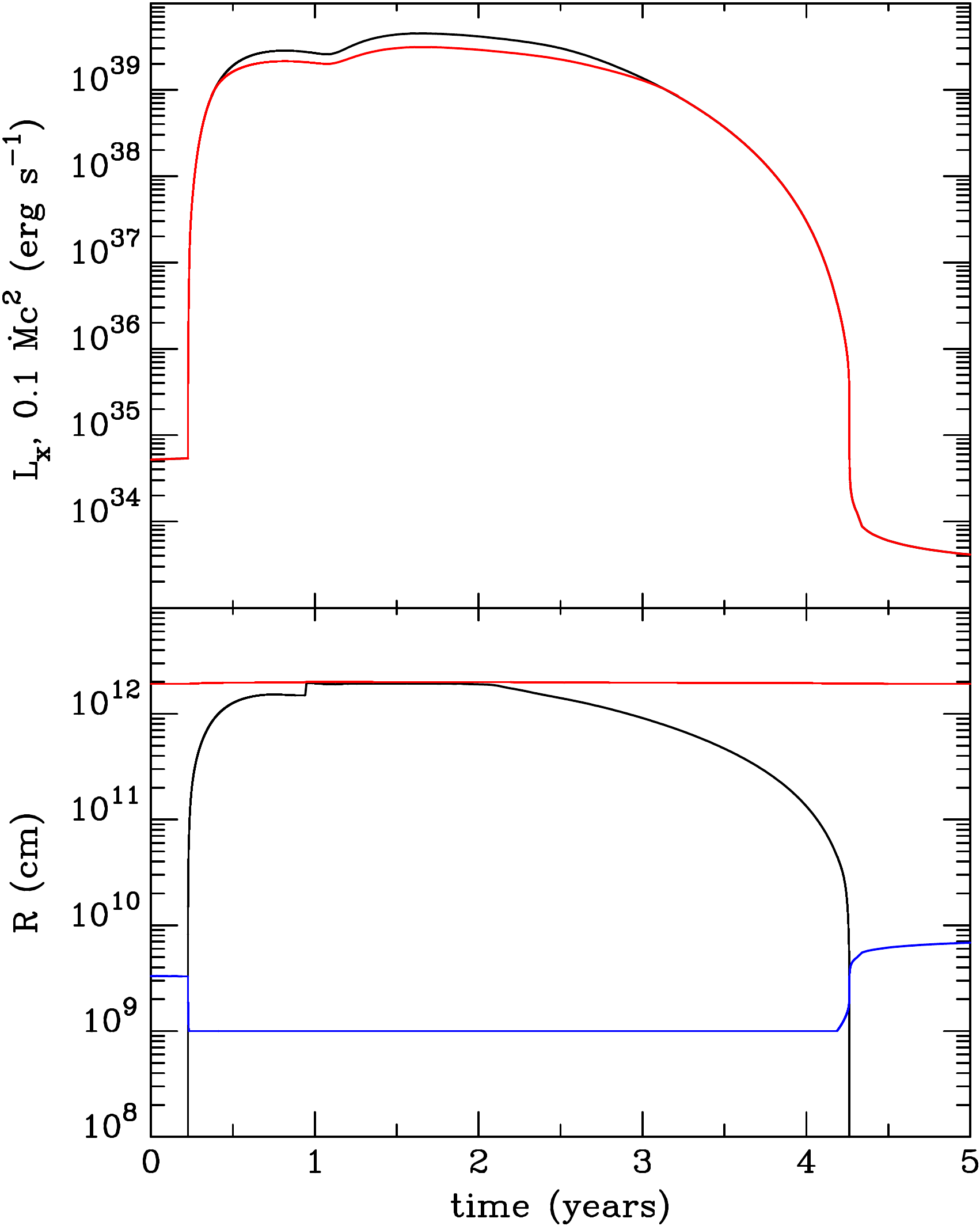}
\caption{Detailed view of a large outburst in Fig. \ref{fig:model12}. The top panel shows the bolometric light curve (red) and the accretion rate times $0.1 c^2$. The bottom panel shows the outer disc radius (red curve), the inner disc radius (blue curve), and the transition radius $r_{\rm tr}$ (black curve).}
\label{fig:model12a}
\end{figure}

\subsection{Outburst sequences}
\label{sec:sequ}

We have calculated the outburst properties for various sets of parameters. Tables 1 and 2 summarize our results in the case where the compact object is a 7~M$_\odot$ black hole (Table 1), or a 1.4~M$_\odot$ neutron star (Table 2). For each model, defined by the mass transfer rate from the secondary $\dot{M}_2$, illumination factor $f_{\rm irr}$, magnetic moment $\mu_{30}$ in units of $10^{30}$~G\,cm$^3$, $\alpha_{\rm c}$ (we kept $\alpha_{\rm h}$ constant and equal to 0.2), we provide the duration of a cycle of outbursts $t_{\rm c}$, defined as the periodicity of the light curve that may cover several outbursts of different intensities, that we classified as large (L; these are the largest outbursts of the sequence), intermediate (m, with peak luminosities ranging between 10 and 100\% of the main outburst values), small (s, 1 to 10\% of the main outburst peak luminosity), and very small ($\mu$, less than 1\% of the main outburst peak luminosity) and for each model, we provide the pattern of the outburst sequence. Tables 1 and 2 also provides the average recurrence time $t_{\rm r}$ between outbursts, that is equal to $t_{\rm c}$ divided by the number of outbursts in the pattern, and, for the largest outbursts, the outburst duration $d$ in years, the maximum accretion rate $\dot{M}_{\rm max}$, the maximum distance reached by the heating front $r_{\rm tr,12}$ in units of 10$^{12}$~cm, the total mass $\Delta M$ accreted during the outburst and $\Delta M/M_{\rm d}$, where $M_{\rm d}$ is the disc mass. 

Models 34 and 34a differ only by the secondary mass, taken arbitrarily to be 5~M$_\odot$ in model 34a, and the orbital period chosen in such a way that the outer disc radius in the same in both cases. Although $r_{\rm circ}$ differs by 40\% between both cases, the outburst properties are, at the percent level, identical. Whether systems with such secondary masses may exist in nature is irrelevant; our point here is merely to show that the outburst characteristics do not depend on $M_2$ for a given outer disc radius.

Figures \ref{fig:model1}, \ref{fig:model12}, and \ref{fig:model12a} show examples of light curve patterns for model 1, a classical case of sub-Eddington outbursts, and 12, a slightly super-Eddington case with $\dot m = 5.5$, at outburst's maximum; its light curve exhibits a complex outburst sequence. 

Figure \ref{fig:model12a} shows that for moderately super-Eddington accretion rates ($\dot m < 8.5$) the difference between the $L$ and $\dot M$ ``light curves'' is negligible for all practical purposes. 

We note that the jump in the transition radius observed when $r_{\rm tr}$ reaches the outer disc radius is due to the fact that, because of tidal heating, the outer disc edge is already in the hot state shortly before the heating front reaches it.

\begin{figure}
\includegraphics[width=\columnwidth]{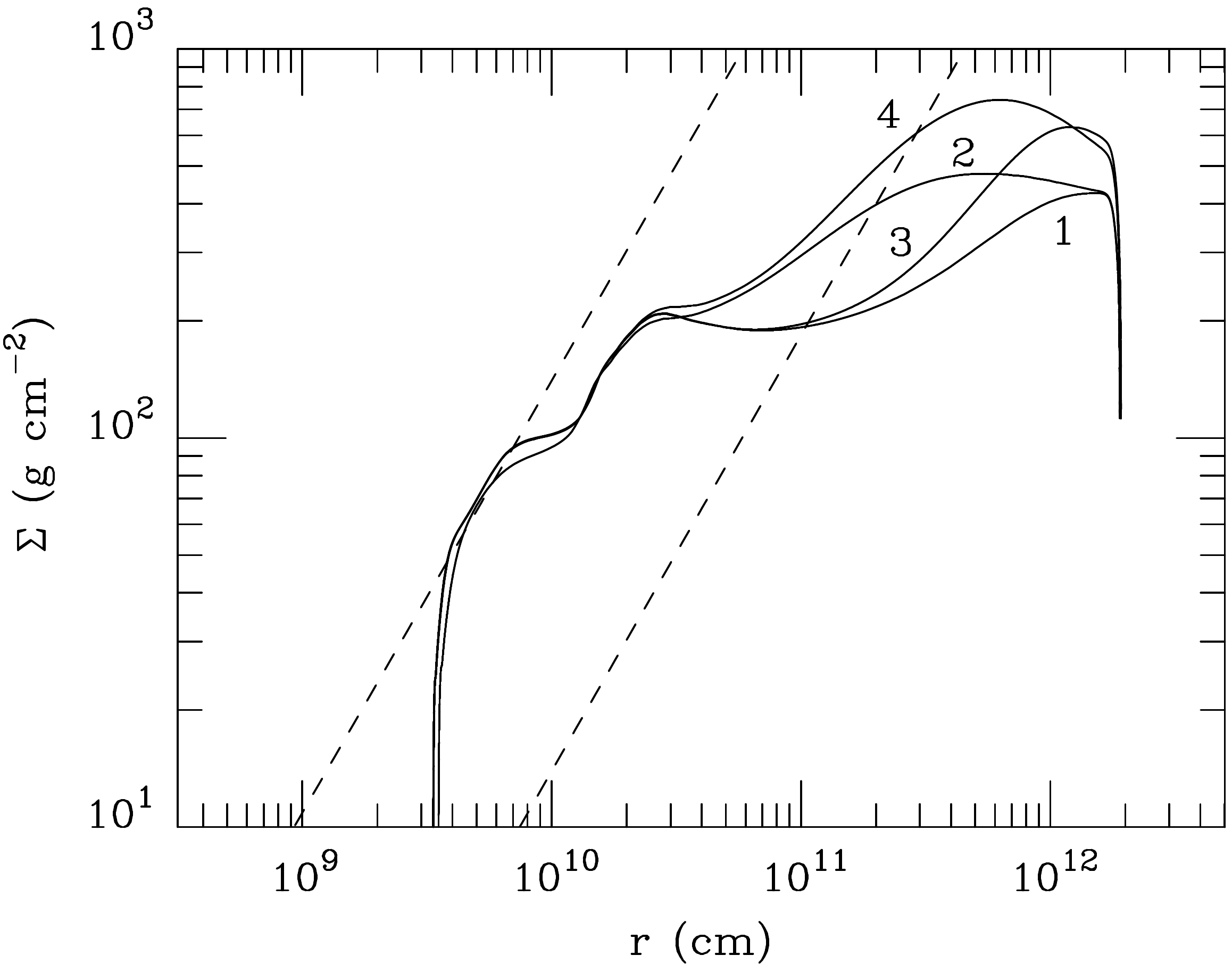}
\caption{Radial surface-density profiles of the accretion disc just prior to outbursts for model 12 (solid lines). The curve labelled 1 shows $\Sigma$ prior to the first weak outburst after a major one, curve 2 shows $\Sigma$ before the second outburst, curve 3 is just before the last weak outburst of the sequence, and curve 4 represents $\Sigma$ at the onset of the last outburst. Dashed lines show the critical surface densities $\Sigma_{\rm min}$ and $\Sigma_{\rm max}$ of the S-curves.}
\label{fig:Sigma-r}
\end{figure}

Figure \ref{fig:Sigma-r} explains why a complex sequence of outbursts can be found in large discs. After a major outburst, the disc density is low; after a time of the order of the diffusion time in the inner parts of the disc, $\Sigma$ reaches the critical density for which the disc can no longer remain in a cold quasi-steady state, and an inside-out outburst is triggered. Because $\Sigma$  in the outer disc is still low, the heating wave cannot propagate very far in the disc, and the outburst amplitude is small. The outburst amplitude, however, has been sufficient to significantly modify $\Sigma$ in intermediate regions of the disc, so that, during the next outburst, the heating front is able to propagate to larger distances, which empties these intermediate regions; during the following outburst, the heating front stops at a smaller distance than for outburst number 2. The sequence continues, and the mass builds up in the outer disc as a result of mass transfer. Eventually, the heating front reaches the outer disc edge, triggering a major outburst. Such a complex sequence of outbursts is reminiscent of the sequences found when modelling discs around AGNs \citep{hvl09}, in which heating fronts are never able to reach the outer edge of the disc.

\begin{figure}
\includegraphics[width=\columnwidth]{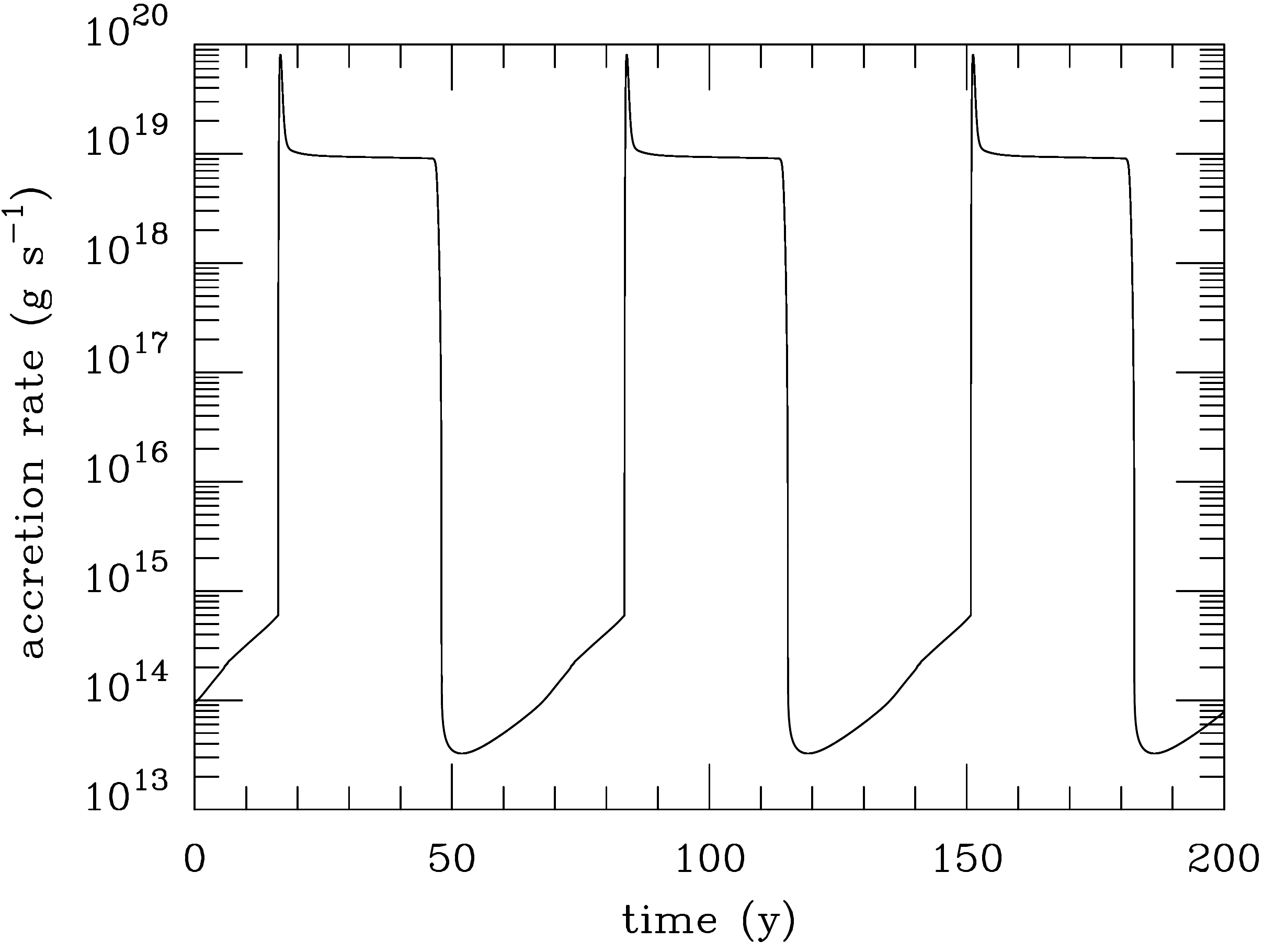}
\caption{Time evolution for model 24, in which the irradiation efficiency $f_{\rm irr}$ decreases when the accretion rate reaches the Eddington value.}
\label{fig:model24}
\end{figure}

\subsection{Impact of disc irradiation}

In all models, except model 24, we have used a constant value of $f_{\rm irr}$.  Figure \ref{fig:model24}, corresponding to model 24, illustrates what happens when in the black-hole case irradiation is Eddington-limited by setting $f_{\rm irr}$ to be
\begin{equation}
f_{\rm irr} = \frac{0.1 \mathcal{C}}{5 \times 10^{-4}} \min \left( 1,\frac{\dot M_{Edd}}{\dot{M}} \right).
\end{equation}
As can be seen, after an initial rapid decay, the accretion rate remain approximately constant at a level slightly above the Eddington value. During this plateau, irradiation remains constant, which prevents the cooling front from propagating inwards; the whole disc, even in the cool outer regions, is close to steady state, and evolves slowly on a characteristic time scale $M_{\rm d}/\dot{M}$ until $\dot{M}$ falls below the Eddington value.

The long plateau observed for model 24 is somewhat reminiscent of the one observed in XTE J1550-564 \citep{smr00}, and is due to constant irradiation temperature at a given disc radius when the accretion rate exceeds the Eddington value, that keeps constant the position of the cooling front. Model 24a corresponds to the same parameters as model 24, with the difference that $\mathcal{C}$ is now assumed to be constant, so that $f_{\rm irr}$ varies as $1 + \ln \dot m$ for $\dot m > 1$. Outbursts now have the usual shape; they are weaker and shorter than in the case where $f_{\rm irr}$ is constant for all luminosities.

For completeness, we have also calculated two models in which irradiation is not taken into account (models 4 and 5). As mentioned above, the recurrence time is short and the total outburst fluence is also much lower than in the irradiated case, even though the peak outburst luminosity is comparable to the irradiated case, leading to a short outburst duration. One should also note that outbursts of non-irradiated discs are terminated by a sequence of reflares, as shown in \citet{dhl01} for smaller accretion discs.

Note that in all our neutron-star models, the accretion rate at maximum is super-Eddington ($\dot{M}_{\rm Edd} = 1.8 \times 10^{18} \; \rm g \, s^{-1}$) and it is also the case of fourteen (out of twenty-four) black-hole models. The mean mass transfer rate is only slightly super-Eddington in models 23, 27, and 35; it is sub-Eddington for all other models, so that none of them has a mass transfer rate usually envisaged for ``steady" ULXs.

The main differences between black hole and neutron star systems are that, for a given orbital period, the disc around a neutron-star binary is smaller, and the Eddington luminosity is lower, which implies that the disc might be more prone to mass loss via a wind for example.

From tables 1 and 2, one can also note that the characteristics of an outbursts do not depend on the disc size, as long as the heating front does not reach the outer disc edge. This is not  surprising as the outer disc regions are not affected by the propagation of heating and cooling fronts and play not role in the outburst outcome.

\subsection{Recurrence time}
\label{sec:rectimes}

Since the condition for the occurrence of outside-in outbursts is  \citep{l01}
\begin{equation}
\dot{M}_2 \gtrsim 4.0\times 10^{20} \delta^{-1 / 2} M_1^{-0.88} 
 \left(\frac{r_{\text {out}}}{10^{12} \mathrm{cm}}\right)^{2.65} \mathrm{g}\, \mathrm{s}^{-1},
\end{equation}
where $\delta \leq 2$ and $r_{\rm out}$ is the outer disc radius in quiescence, all models considered here correspond to inside-out propagation of the heating front, i.e., to outbursts starting near the inner disc's edge.
\begin{figure}
\includegraphics[width=\columnwidth]{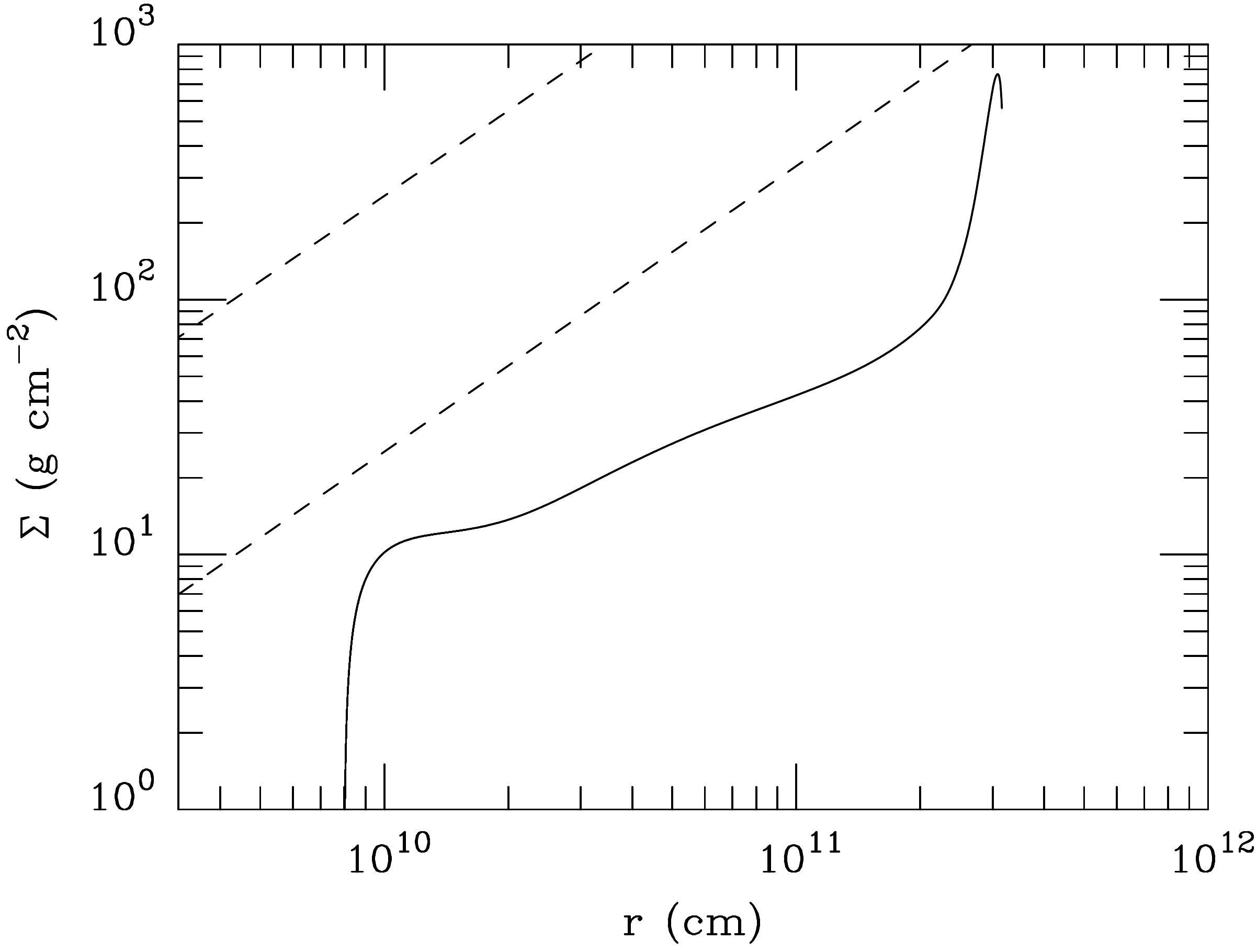}
\caption{Quiescent radial surface-density profile 173 days after the end of an outburst for model 26. The dashed lines correspond to critical surface densities $\Sigma_{\rm max}$ (upper line) and $\Sigma_{\rm min}$ (lower line).}
\label{fig:profil25}
\end{figure}
For such outbursts, the recurrence time is approximately equal to the viscous time at the innermost parts of the accretion disc, and should thus depend only on $\alpha_{\rm c}$ and on the inner disc radius in quiescence. Tables 1 and 2 shows that the recurrence time is the same for all models that share the same $\alpha_{\rm c}$, with the exception of models 4, 5, and 24, and is almost independent of the mass transfer rate. In models 4 and 5, the recurrence time is short because, neglecting irradiation causes reflares that alter the surface density profile as compared to the case where a cooling front is able to propagate directly to the disc inner edge \citep{mhln2000,dhl01}. In model 24, the inner disc is largely depleted after long outbursts and slow decay to quiescence (see below); the inner disc radius is larger than in other models, leading to a long recurrence time.

The recurrence time of outbursts in X-ray irradiated discs is longer than in dwarf novae (except for WZ Sge-type ``super-outbursts''), and is not given by Eq. 53 in \citet{l01} which applies to systems in which outer-disc irradiation is negligible (such as dwarf novae). In such a case the quiescent $\Sigma$ is in the range $[\Sigma_{\rm min},\Sigma_{\rm max}]$, where $\Sigma_{\rm min}$ and $\Sigma_{\rm max}$ are the turning points of the S curve in the $\Sigma$ -- temperature diagram. As explained  in \citet{l01}, this is not longer true when, due to irradiation, the disc, and in particular its inner portions, are strongly depleted during outbursts.  This is illustrated by Fig. \ref{fig:profil25} which shows the $\Sigma(r)$ profile at the end of an outburst; $\Sigma$ is much below the non-irradiated (irradiation plays no role in quiescence) $\Sigma_{\rm min}$, which explains the long recurrence time in these models: at the start of quiescence, quiescent discs can be almost empty. As seen in Tables 1 and 2, up to 90\% of the disc mass can be accreted during outbursts.

The recurrence time can be also estimated from the disc refilling time
\begin{equation}
t_{\mathrm{refill}}  = \frac{\Delta M}{\dot{M}_{2}} = 31.7 \left(\frac{\Delta M}{10^{27}\; \rm g}\right)\left(\frac{\dot M_2}{10^{18}\rm \; g \, s^{-1}}\right)^{-1} \rm yr,
\end{equation}
in agreement with the $t_{\mathrm{r}}$ values given in Tables 1 and 2.

Notice that, in contrast with the dwarf nova case, the disc mass in our transients is never close to maximum (corresponding to $\Sigma \sim \Sigma_{\rm max}$) as clearly seen in Fig. \ref{fig:Sigma-r}.

This said, one should keep in mind that our standard assumption that in quiescence the disc accretion is driven by a process describable by an $\alpha$--prescription is far from being guaranteed. As studied in detail by \citet{sc2018}, in accretion discs with $T \lesssim 3500$~K and $\Sigma \lesssim 180$~g\,cm$^{-2}$, the magnetic Reynolds number
$R_{\mathrm{m}} = {c_{\mathrm{s}} H}/{\eta_{\rm R}}$, where $c_{\rm s}$ is the sound speed, $H$ the disc's pressure scale-height and $\eta_{\rm R}$ the Ohmic resistivity coefficient, is less than $10^4$ so that diffusion of the magnetic field becomes too important for the disc to sustain the MHD turbulence \citep[][see also \citet{g1998}]{h1996,f2000}. Observed quiescent X-ray fluxes prove that accretion is occurring also during this phase of the outburst cycle \citep[see eg.,][]{l2000} but since in quiescent X-ray transient discs $T$ is less than 3500~K and $\Sigma$ less than 180~g\,cm$^{-2}$ (see for example Figure \ref{fig:profil25}) it is not known what drives accretion in quiescent XBTs. The calculated values of recurrence times should be therefore subject to caution, even if they are in good agreement with observations \citep[compare for example with][]{c2012}.

\section{Analytical relations} \label{sect:analytic}

\subsection{Peak accretion rate}

As noted by \citet{lkd15}, there exists a tight correlation between the maximum accretion rate during an outburst $\dot{M}_{\rm max}$ and the maximum distance reached by the transition front $r_{\rm tr,max}$,  because at the outburst peak, the portion of the disc that is in the hot state is close to being steady, and the mass accretion rate is thus equal to the minimum (critical) rate $\dot{M}_{\rm crit}^+$ for which a hot stable disc can still exist. This critical rate is well approximated by:
\begin{equation}
    \dot{M}_{\rm crit}^+ \approx 2.4 \times 10^{19} M_1^{-0.4}f_{\rm irr}^{-0.5} \left( \frac{r_{\rm tr,max}}{10^{12} \; \rm cm} \right)^{2.1} \; \rm g \, s^{-1}
\label{eq:mdotcrit}
\end{equation}
Here we use the formulae for critical quantities from \citet{l01} that are slightly different from those in \citet{lkd15} who used fits obtained in \citet{ldk08}. The version of the DIM code used in this paper differs in some minor aspects from that used in this reference.

\begin{figure}
\includegraphics[width=\columnwidth]{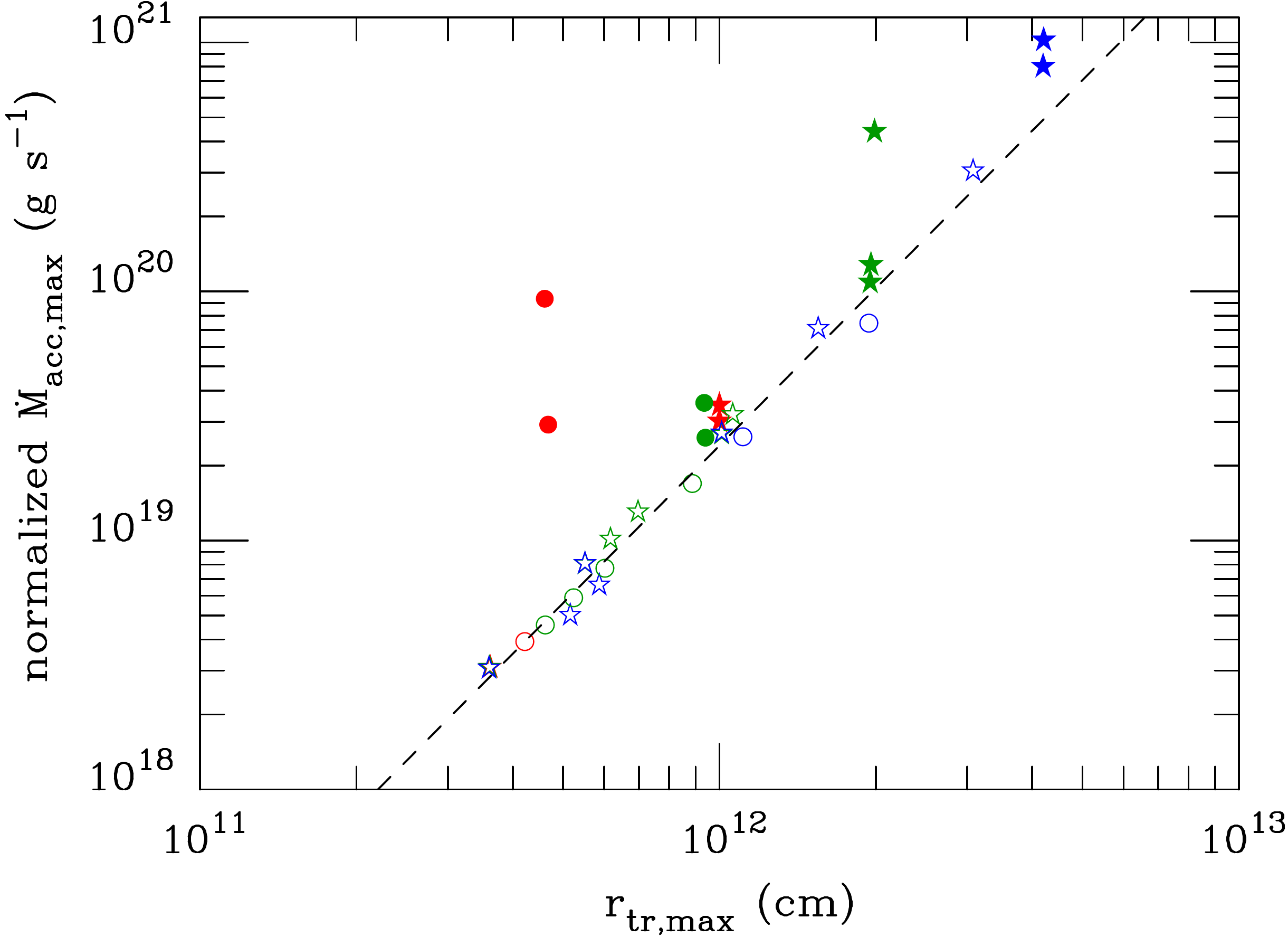}
\caption{Relation between the maximum mass accretion rate reached during an outburst multiplied by $M_1^{0.4} f_{\rm irr}^{0.5}$ and the maximum distance travelled by the heating front. Stars correspond to black hole systems, circles to neutron star systems. Red symbols correspond to an orbital period of 155~hr, green symbols to 400~hr and blue symbols to 1200~hr. Symbols are filled when the heating front reached the outer edge of the disc. The dashed line represents the critical mass transfer rate as given by Eq. \ref{eq:mdotcrit}. }
\label{fig:Mdot-rtr}
\end{figure}

Figure \ref{fig:Mdot-rtr} shows all our models in the plane $(r_{\rm tr,max},\dot{M}_{\rm max})$ where $\dot{M}_{\rm max}$ has been normalized by $M_1^{0.4} f_{\rm irr}^{0.5}$.  We have omitted models 4 and 5, in which irradiation is not taken into account, as well as model 24 in which $f_{\rm irr}$ varies. 

As can be seen, the agreement between the maximum accretion rate during an outburst and $\dot{M}_{\rm crit}^+$ is remarkable as long as the heating front does not reach the outer-disc edge. Moreover, if we use the fits for $\dot{M}_{\rm crit}^+$ appropriate for the unirradiated case given in \citet{bhl18}, we find that the peak accretion rate for model 4 is 1.36 times the critical rate at the transition radius, and this ratio is 1.07 for model 5, so that this agreement also holds for the two models that are not shown in Fig. \ref{fig:Mdot-rtr}.

If the heating front reaches the disc's outer edge, the peak accretion rate is larger than what is predicted by Eq. \ref{eq:mdotcrit}. This corresponds to filled symbols in Fig. \ref{fig:Mdot-rtr}. $\dot{M}_{\rm max}$ can reach values higher than predicted by up to a factor 50. The higher $\dot{M}_2$, the higher $\dot{M}_{\rm \max}$, with a limit due to the fact that $\dot{M}_2$ cannot be too high in transient systems, as we discuss in the next section. The upper red dot in Fig. \ref{fig:Mdot-rtr} corresponds to model 27 for which $\dot M_2=2\times 10^{18}$g/s is rather close to the stability limit ($4.9\times 10^{18}$g/s). This does not prevent, however, the disc to exhibit full-fledged outbursts.

It is worth noting that the $r_{\rm tr,max}$ -- $\dot{M}_{\rm max}$ relation also approximately holds to within a factor two for the actual accretion rate during an outburst, with $\dot{M}$ slightly below the fit during the rise and slightly above during decay. We use this property in Sect. \ref{sect:coolingfront} to derive the propagation time of cooling fronts.

The heating front can reach the outer-disc edge if the mass transfer rate is high enough. A close look at table 1 shows that, to first order, $\dot{M}_{\rm max}/\dot{M}_2$ depends  mainly on the ratio $\alpha_{\rm h}/\alpha_{\rm c}$, provided that the heating front does not reach the outer disc edge. $\dot{M}_{\rm max}/\dot{M}_2$ is of the order of 25 to 35  when $\alpha_{\rm h}/\alpha_{\rm c} = 5$, and of the order of 70 to 100 when $\alpha_{\rm h}/\alpha_{\rm c} = 10$. Deviations are observed when there is a long tail in the outburst, as in the case of Eddington limited irradiation; the low $\dot{M}_{\rm max}/\dot{M}_2$ found in this case results from the depletion of the accretion disc during the plateau that prevents the heating front to propagate at large distances during the next outburst. It also appears that $f_{\rm irr}$ has a limited impact on $\dot{M}_{\rm max}/\dot{M}_2$; a decrease in irradiation changes the relation between $r_{\rm tr}$ and $\dot{M}_{\rm max}$, but not $\dot{M}_{\rm max}$ itself. The relation
\begin{equation}
    \dot{M}_{\rm max} = \phi(\alpha_{\rm h}/\alpha_{\rm c}) \dot{M}_2
\label{eq:mdotmax}
\end{equation}
also holds when the transition front reaches the outer disk edge, but with some deviation that increases with increasing $\dot{M}_2$, in particular when the outburst duration is no longer much shorter than the recurrence time.

Using this approximate relation together with $\dot{M}_{\rm max} = \dot{M}_{\rm crit}^+$ enables to determine $r_{\rm tr,max}$. If the latter quantity is larger than the outer disc radius $r_{\rm out}$, the heating front reaches the outer disc edge, and the full disk is brought in the hot state. In the opposite case, a cooling front is initiated at $r_{\rm tr,max}$.

\subsection{Outburst decay time and duration}

During outburst maximum and decay, the hot disc region is close to being steady \citep{l01}. The surface density is then
\begin{equation}
\Sigma = 66.2\, \psi \alpha_{0.2}^{-4/5} \dot{M}_{19}^{7/10} M_1^{1/4} r_{12}^{-3/4} \; \rm g \, cm^{-2}
\label{eq:sigma}
\end{equation}
where $\alpha_{0.2} = \alpha/0.2$, $\dot{M}_{19} = \dot{M}/10^{19}$~g\,s$^{-1}$, $r_{12} = r/10^{12}$~cm, and $\psi \sim 1$ describes the deviation of the opacities from the Kramers' law used by \citet{l2016} to obtain Eq. \ref{eq:sigma}. Our calculations of the effective temperature as a function of surface density yield values for $\psi$ varying between 1.0 and 1.5; in the following, we use $\psi = 1.3$. If the full disk is hot, its mass is
\begin{equation}
    M_{\rm d} = 3.3 \times 10^{26} \psi \alpha_{0.2}^{-4/5} \dot{M}_{19}^{7/10} M_1^{1/4} r_{12}^{5/4} \; \rm g,
    \label{eq:md}
\end{equation}
The estimates provided by Eqs. \ref{eq:sigma} and \ref{eq:md} require that irradiation does not significantly alter the disc vertical structure, which is the case when the irradiation temperature is significantly lower than the disc central temperature. We have checked that the disc mass obtained using Eq. \ref{eq:md} corresponds to better than a factor 1.5 to the actual calculated disc mass when the heat front does reach the outer disk edge.

\begin{figure}
\includegraphics[width=\columnwidth]{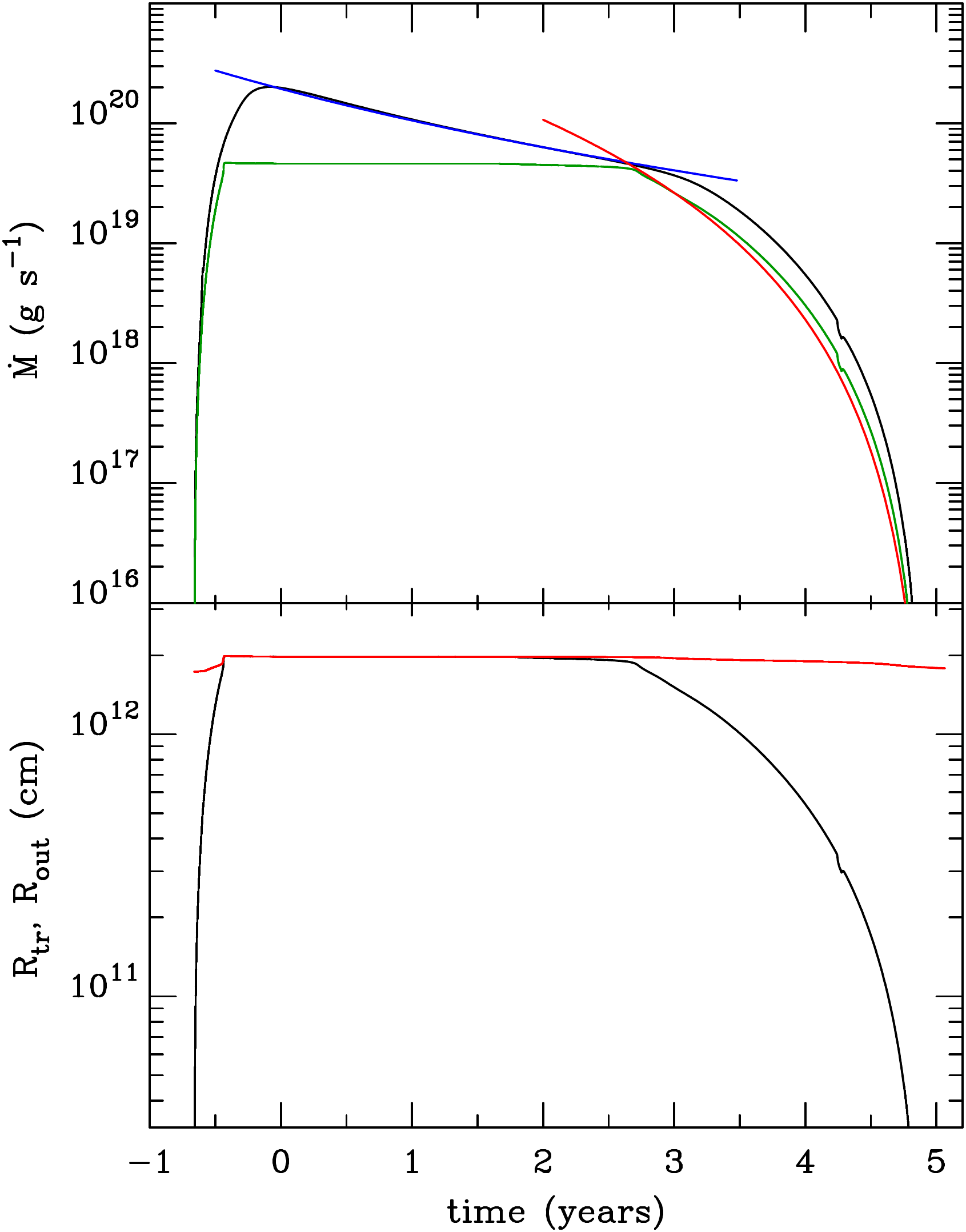}
\caption{Time evolution of model 14. The bottom panel shows the transition radius (black curve) as well as the outer disc radius (red curve). The top panel shows the actual mass accretion rate (black curve) and the critical rate $\dot{M}_{\rm crit}^+$ (green curve). The blue curve gives the analytic solutions (Eq. \ref{eq:varmdot}) when the entire disc is on the hot state, and the red curve corresponds to a propagating cooling front (Eq. \ref{eq:decay}).}
\label{fig:decay}
\end{figure}

\subsubsection{Decay of a fully hot disc}
\label{sec:hot}
We first consider the case where the heat front reaches $r_{\rm out}$. During the initial decay, the disc mass decreases while remaining entirely in the hot state; Eq. \ref{eq:md} relates the disc mass to its time derivative; it can be solved and we obtain \citep{rk01}:
\begin{equation}
\dot{M}=\dot{M}_{\rm max} \left[ 1+ \frac{t}{t_0} \right]^{-10/3},  
\label{eq:varmdot}
\end{equation}
where we corrected a misprint in \citet{rk01} Eq. 10. $t_0$ is given by: 
\begin{equation}
    t_0 = 2.45 \psi \alpha_{\rm 0.2}^{-4/5} M_1^{1/4} r_{12}^{5/4} \dot{M}_{\rm max,19}^{-3/10} \; \rm yr,
\end{equation}
where $\dot{M}_{\rm max,19} = \dot{M}_{\rm max}/10^{19}$~g\,s$^{-1}$. Although related to it, $t_0$ is {\sl not} the characteristic time scale of the disc evolution. We can define a characteristic timescale from Eqs. \ref{eq:md}  and \ref{eq:mdotcrit}
\begin{equation}
    \tau = \frac{M_d}{\dot{M}} = 0.81 \psi f^{-0.3} M_1^{0.37} f_{\rm irr}^{0.15} r_{12}^{0.62} \alpha_{0.2}^{-0.8} \; \rm yr, 
\label{eq:tau}
\end{equation}
where $f=\dot{M}_{\rm max}/\dot{M}_{\rm crit}^+(r_{\rm out}) > 1$. $f$ is given by 
\begin{equation}
   f \sim \phi(\alpha_{\rm h}/\alpha_{\rm c}) \dot{M_2}/ \dot{M}_{\rm crit}^+(r_{\rm out})
\end{equation}
and is less than $\phi$ since the mass transfer rate has to be less than the critical rate for the system to be transient.

Note, that contrary to the common opinion \citep[e.g.][]{kr98,dhl01,lkd15} the light curves of outbursts produced by the instability of irradiated discs in X-ray transients systems are not exponential \citep[see also ][]{k98}.

Figure \ref{fig:decay} compares the time evolution of the mass accretion rate as found in model 14 from our numerical simulations with the analytical estimate given by Eq. \ref{eq:varmdot}. The little wiggle seen at time $t\sim 4.2$~yr is due to the ad hoc truncation of the disc during the outburst maximum for this model, see Sect. \ref{sec:model}. The approximation is quite good and relies on two free parameters: the time at which the decay from maximum starts and the peak mass accretion rate. In the case shown in Fig. \ref{fig:decay}, we use $\dot{M}_{\rm max}=1.94 \times 10^{20}$~g\,s$^{-1}$, and the corresponding $t_0$ is 5.0~yr. As mentioned above, $t_0$ is longer than the characteristic evolution time, which is 2.0~yr. We also stress again that, although the decay is not very different from exponential, a -10/3 power law is by far a much better fit.

We can relate the peak accretion rate to the characteristic decay time by combining Eqs. \ref{eq:mdotcrit} and \ref{eq:tau}; we find:
\begin{equation}
    \dot{M}_{\rm max} = 4.9 \times 10^{19} \alpha_{0.2}^{2.71} f^{2.02} M_1^{-1.65} \left( \frac{\psi^{-1} \tau}{1 \; \rm yr}\right)^{3.39} f_{\rm irr}^{-1.01}  \; \rm g \, s^{-1}.
\label{eq:mdot-tau}
\end{equation}
If, from observations, both $\dot{M}_{\rm max}$ and $\tau$ are known, this relation determines $f$, and hence $\dot{M}_{\rm crit}^+$ and the size of the accretion disc.

A slightly different approach was used by \citet{lkd15} who assumed that the decay from maximum of a fully hot, irradiated disc begins with an approximately exponential phase, during which matter is accreted at a viscous time at the constant (during this phase) outer  disc radius \citep{dhl01,kr98}, the characteristic viscous time being given by $t \sim r^{2}/(3\nu)$.

Therefore the decay time is equal to the viscous time and can be written as
\begin{equation}
\tau_1 \simeq \frac{(G M r)^{1 / 2}}{3 \alpha c_{s}^{2}},
\end{equation}
where $c_s=kT_c/m_{\rm p}$, with $T_c\approx T_{\rm crit}^+$ the critical minimal midplane temperature of the hot irradiated disc. Taking $T_c \approx 27200 \, (r_{\rm out}/10^{12}\,\rm cm)^{0.05}$\,K \citep{l01}, we get:
\begin{equation}
\tau_{1} \approx 0.49 \,M_1^{1 / 2}\,  \,\alpha_{0.2}^{-1} \left( \frac{r}{10^{12} \rm cm}\right)^{0.45}\text {yr}.
\end{equation}
This is close to the decay time found in Eq. \ref{eq:tau}, although the parameter dependence is rather different. This difference strongly affects the $\dot M_{\rm max} (\tau)$ relation and explains why our Eq. \ref{eq:mdot-tau} and Eq. 6 of \citet{lkd15} are so dissimilar.

\subsubsection{Propagation of a cooling front} \label{sect:coolingfront}
When $\dot{M}$ falls below $\dot{M}_{\rm crit}(r_{\rm out})$, a cooling front starts propagating from the outer-disc edge. The situation is similar if the heating front does not reach the outer disc edge but in this latter case, the cooling fronts starts inside the disc. We can then use the same method for determining the time evolution of the disc, the main differences being that the disc mass in Eq. \ref{eq:md} now refers to the mass of the fraction of the disc that is in the hot state, and $\dot{M}$ is no longer the time derivative of the hot disc mass. Instead we can write
\begin{equation}
\dot{M}_{\rm d} = -\dot{M} - \dot{M}_{\rm tr}+ 2 \pi r_{\rm tr} \Sigma \dot{r}_{\rm tr}, 
\end{equation}
where the dots indicate time derivatives and $\dot{M}_{\rm tr}$ is the mass flow at the transition radius\footnote{{\it not} the mass transfer rate.}. Our simulations indicate that $\dot{M}_{\rm tr}$ is positive (i.e. mass flows outwards) and comparable to $\dot{M}$ \citep[see][for a detailed discussion of the structure of the transition fronts]{mhs99,dhl01}. Since $\dot{M}=\dot{M}_{\rm crit}^+(r_{\rm tr})$, we can express $r_{\rm tr}$ as a function of $\dot{M}$ from Eq. \ref{eq:mdotcrit}, and after some algebra, we get
\begin{equation}
\dot{M}_{\rm d} = -2.47 (\dot{M} + \dot{M}_{\rm tr}) =  \xi \dot{M}.  
\label{eq:mdot1}
\end{equation}
If we make the additional assumption that $\xi$ does not vary much with time, we can solve Eq. \ref{eq:md} using Eqs. \ref{eq:mdotcrit} and \ref{eq:mdot1}; we find:
\begin{equation}
    \dot{M} = 3.19 \times 10^{17} \alpha_{0.2}^{2.71} M_1^{-1.65} f_{\rm irr}^{-1} \left[ \frac{\xi}{\psi} \frac{(t_{0}^{\prime}-t)}{1 \; \rm yr} \right]^{3.39} \; \rm g \, s^{-1},
    \label{eq:decay}
\end{equation}
where $t_0^{\prime}$ is a constant that is determined by the condition that, when the cooling front starts, $\dot{M}$ is equal to $\dot{M}_{\rm crit}^+$ at the maximum transition radius. $t_0^\prime$ can then be written as:
\begin{equation}
    t_0^{\prime}=3.57 \; \psi \xi^{-1} M_1^{0.37} f_{\rm irr}^{0.15} \alpha_{0.2}^{-0.8} r_{12}^{0.62} \; \rm yr. 
    \label{eq:t0}
\end{equation}

\begin{figure}
\includegraphics[width=\columnwidth]{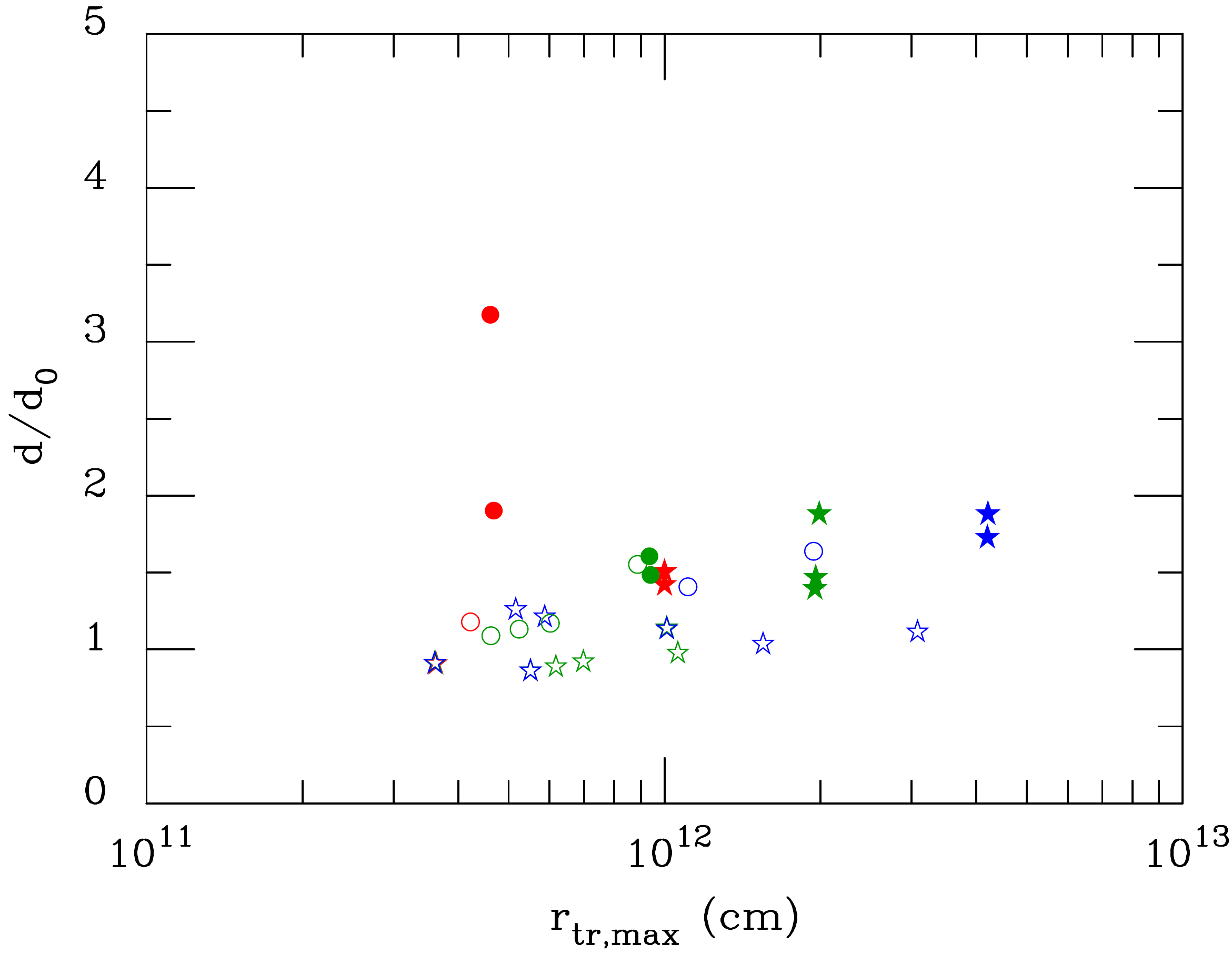}
\caption{Comparison between the outburst duration and the duration of the propagation time of the cooling wave as estimated by Eq. \ref{eq:decay} with $\xi = 5$. The symbols have the same meaning as in Fig. \ref{fig:Mdot-rtr}.}
\label{fig:d-rtr}
\end{figure}

Figure \ref{fig:decay} compares the time evolution of the accretion rate as determined by simulations for model 14 with the analytical approximation above. Here, there is only one free parameter, $\xi$, and we took $\xi = 6.3$. As can be seen, Eq. \ref{eq:decay} represents quite well the results of numerical simulations. Interestingly, the differences are mainly due to the fact that the accretion rate is not exactly equal to the critical rate at the transition radius; Eq. \ref{eq:decay} fits very well the critical rate $\dot{M}_{\rm crit}^+$, much better than the actual accretion rate $\dot{M}$.

It must also be stressed out that $t_0^{\prime}$ is equal to the duration of this phase, and it can thus be directly compared to observable quantities. Figure \ref{fig:d-rtr} shows the ratio of the duration $d$ of the outburst divided by $d_0=t_0^{\prime}$, for $\xi=5$, corresponding to $\dot{M}_{\rm tr}=\dot{M}$. Because the rise from quiescence is much shorter than the decay, $d$ is a good estimate of the duration of the final decay phase. When the heating front does not reach the outer disc edge, $d$ is close to $t_0^{\prime}$, to better than a factor 1.5. This is not the case when the full disk is brought to a hot state, because there exists a phase during which the disc decays while being entirely in the hot state.

It is interesting to note that $t_0^\prime$ and $\tau$, as given by Eqs \ref{eq:t0} and \ref{eq:tau}, have the same functional dependence and are of the same order. This is not a surprise, since both equations results from Eqs. \ref{eq:mdotcrit} and \ref{eq:md}. As a consequence, the relation between $\dot{M}_{\rm max}$ and $\tau$ is essentially the same as in Eq. \ref{eq:mdot-tau}, with $f=1$; it writes:
\begin{equation}
    \dot{M}_{\rm max} = 7.6 \times 10^{19} \alpha_{0.2}^{2.71} M_1^{-1.65}  \times \left( \psi^{-1} \frac{\xi}{5}\frac{t_0^\prime}{1 \; \rm yr}\right)^{3.39} f_{\rm irr}^{-1.01}  \; \rm g \, s^{-1}.
\label{eq:mdot-tp0}
\end{equation}

As a note of caution, these results have been obtained under the assumption that $\xi$ is constant. There is no reason to believe that this is the case, nor that $\xi$ should not depend on parameters such as $M_1$, $C$ or $\alpha$. The goodness of the fits seems to indicate, however, that $\xi$ is not a sensitive function of these parameters. One reason for this is that, by definition, $\xi > 2.47$; and, from our experience, $\dot{M}_{\rm tr}$ is never much bigger than $\dot{M}$.

\section{Comparison with observations} \label{sect:obs}

As has been known for some time, the irradiated-DIM accounts well for the observations of classical galactic transients \citep{c2012} and provides basic information about their observed light-curves \citep{dhl01} but requires refinement if it is to reproduce really observed outbursts \citep{tdl18,tlh18}. 

For accretion rates less than the Eddington value, the luminosity is proportional to $\dot m$ and the decay time of the luminosity $\tau_{\rm L}$ is equal to the decay time of the accretion rate. From Eq. \ref{eq:mdot-tau}, in order to have $\dot m_{\rm max} \leq 1$ we need
\begin{equation}
    \tau_{\rm L} \leq 0.34\, \psi\, \alpha_{0.2}^{-0.8} f^{-0.6} M_1^{0.78} f_{\rm irr}^{0.3} \, \rm yr.
\label{eq:taul}
\end{equation}
This (taking $\psi=1.3$) is in reasonable agreement with Eq. 10 in \citet{lkd15} who used $f=3$, and found a critical $\tau_{\rm L}$ smaller by a factor of about 1.3. Eq. \ref{eq:taul} is, however, more general since it contains explicitly $f$ which is implicitly related to the disc size at maximum.

If, on the other hand, $\dot m \gg 1$ so that the apparent luminosity scales approximately as $\dot m^2$, the decay time of the luminosity $\tau_{\rm L}$ is half the decay time of the accretion rate, and Eq. \ref{eq:mdot-tau} can be written as:
\begin{equation}
    \dot{m}_{\rm max} = 395\, \alpha_{0.2}^{2.71} f^{2.02} M_1^{-2.65}
      \left( \frac{\psi^{-1} \tau_{\rm L}}{1 \; \rm yr}\right)^{3.39} f_{\rm irr}^{-1.01}.
\label{eq:mdot-tau_edd}
\end{equation}
The comparison with Eq. 17 of \citet{lkd15} is a bit pointless because in both formulae the parameters are raised to rather high but somewhat different powers.

If the entire disc is not brought in the hot state, the total outburst duration is related to the maximum mass accretion rate via Eq. \ref{eq:mdot-tp0}, and we must have:
\begin{equation}
    \dot{m}_{\rm max} = 58.5\, \alpha_{0.2}^{2.71} M_1^{-2.65}
      \left( \psi^{-1} \frac{\xi}{5}\frac{t_0^\prime}{1 \; \rm yr}\right)^{3.39}
      f_{\rm irr}^{-1.01},
\label{eq:mdot-tp0_edd}
\end{equation}
where $t_0^\prime$ now refers to the outburst duration, longer than the decay time that enters Eq. \ref{eq:mdot-tau_edd}. Equation \ref{eq:mdot-tp0_edd} does not apply when the entire disc is brought in the hot state, because in this case the outburst duration is longer than the propagation time of a cooling wave and because $\dot m_{\rm max}$ is higher than the accretion rate for which a cooling wave starts by a factor $f$, possibly much larger than unity.

These equations confirm the main argument of the present paper, i.e., that to study super-Eddington outbursts one has to consider large discs. A detailed confrontation of Eqs. \ref{eq:mdot-tau_edd} and \ref{eq:mdot-tp0_edd} with reality could be attempted and is encouraged, keeping in mind that this has been done already for sub-Eddington outbursts suggesting that $\alpha_h > 0.2$ \citep{tlh18} and that $f_{\rm irr}$ is constant neither in time, nor in space, with possible values larger than 1 \citep{tdl18}. These papers, however, used the \citet{kr98} exponential-decay formalism so that checking the influence of this assumption on the conclusions about $\alpha_{\rm h}$ and $f_{\rm irr}$ is probably worth trying. Also Eq. \ref{eq:mdot-tau} could be used to check the consistency of the ``observed''  $\alpha_{\rm h}$ and $f_{\rm irr}$ with the disc size. In principle, for transient ULXs, as for LMXBs, it should be possible, to determine from observations if a phase during the entire disc is brought to a hot state exists, because there should be a change of slope in the light curve; the existence of such a break might, however, be difficult to assess if the signal to noise ratio is not high enough.

This said, it is worth using the numerical models directly to compare them with the observations of X-ray outbursts with well sampled and not too extravagant light curves which allow to test the credibility of the values of the assumed parameters. We discuss first the case of V404 Cyg, a galactic transient source with an orbital period of 155~hr, before considering more extreme cases, in terms of luminosity and possibly of disc size.

\subsection{The case of V404 Cyg} 
\label{sect:v404}
The parameters of model 1 are adequate for a system such as V404 Cyg, i.e. $M_1$=7~M$_\odot$ and $P_{\rm orb}= 155$~h. The orbital period of V404 Cyg is about 6.5 ~d \citep{ccn92}, and the mass function is $f(M)=6.1$~M$_\odot$, with a mass ratio $M_2/M_1=0.060^{+0.004}_{-0.005}$ \citep{cc94}. The secondary is a K2--4 giant \citep{kfr10}, located at a distance of $2.39\pm0.14$~kpc \citep{mjd09}, in agreement with the {\it GAIA} parallax of $0.44 \pm 0.10$~mas \citep{grj19}. Integrating the observed X-ray light curve, \citet{zds99} estimated that the total accreted mass during the 1989 outburst was $6 \times 10^{25}$~g, assuming a radiative efficiency of 0.1 and a distance of 3.5 kpc; rescaling to the currently accepted distance of 2.39 kpc leads to an estimate of the mass transferred during this outburst of $2.6 \times 10^{25}$~g \citep{zz18}, implying an average accretion rate of the order of $2.5 \times 10^{16}$~g\,s$^{-1}$ if the amount of mass accreted in quiescence is negligible.

Figure \ref{fig:model1} shows the time evolution of the accretion disc for parameters that are compatible with those of V404 Cyg: the mass transfer rate from the secondary is $5\times 10^{16}$~g\,s$^{-1}$; we used $\alpha_{\rm h} = 0.2$ and $\alpha_{\rm c} = 0.04$, and took $\eta_{\rm t} \mathcal{C}=5 \times 10^{-4}$, as in \citet{dhl01}. We used $\mu = 10^{31}$~G\,cm$^3$. We find a sequence of regular outbursts, lasting 11 months, with a peak accretion rate of $1.4 \times 10^{18}$~g\,s$^{-1}$, and recurring every 12 years. These characteristics compare reasonably well with the observed properties of V404 Cyg which showed three outbursts in 1938, 1989, and 2015, implying a recurrence time of the order of 30 -- 50~yr. The recurrence time we obtain is shorter than that observed in V404~Cyg; longer recurrence times can easily be obtained by reducing the viscosity in the cold state $\alpha_{\rm c}$, as we have seen in section \ref{sec:rectimes}. The peak luminosity in 1989 was about $8 \times 10^{38}$~erg\,s$^{-1}$, and the outburst lasted for about 400~d. The 2015 outburst was very  different from the 1938 and 1989 events \citep[see e.g.][]{cmm19}. These authors suggest that the outburst has been terminated by a very strong outflow from the outer-disc regions and invoke the possibility of an enhanced mass transfer during the outburst. As discussed above, winds are not properly taken into account in the standard DIM. Enhanced mass transfer is routinely added to the DIM description of dwarf nova outbursts \citep[see e.g.][]{Hameury19}, but only rarely invoked in the case of X-ray transient sources \citep[see][for an example]{elh2000}.

\subsection{Ultra-luminous X-ray sources}

\begin{figure}
\includegraphics[width=\columnwidth]{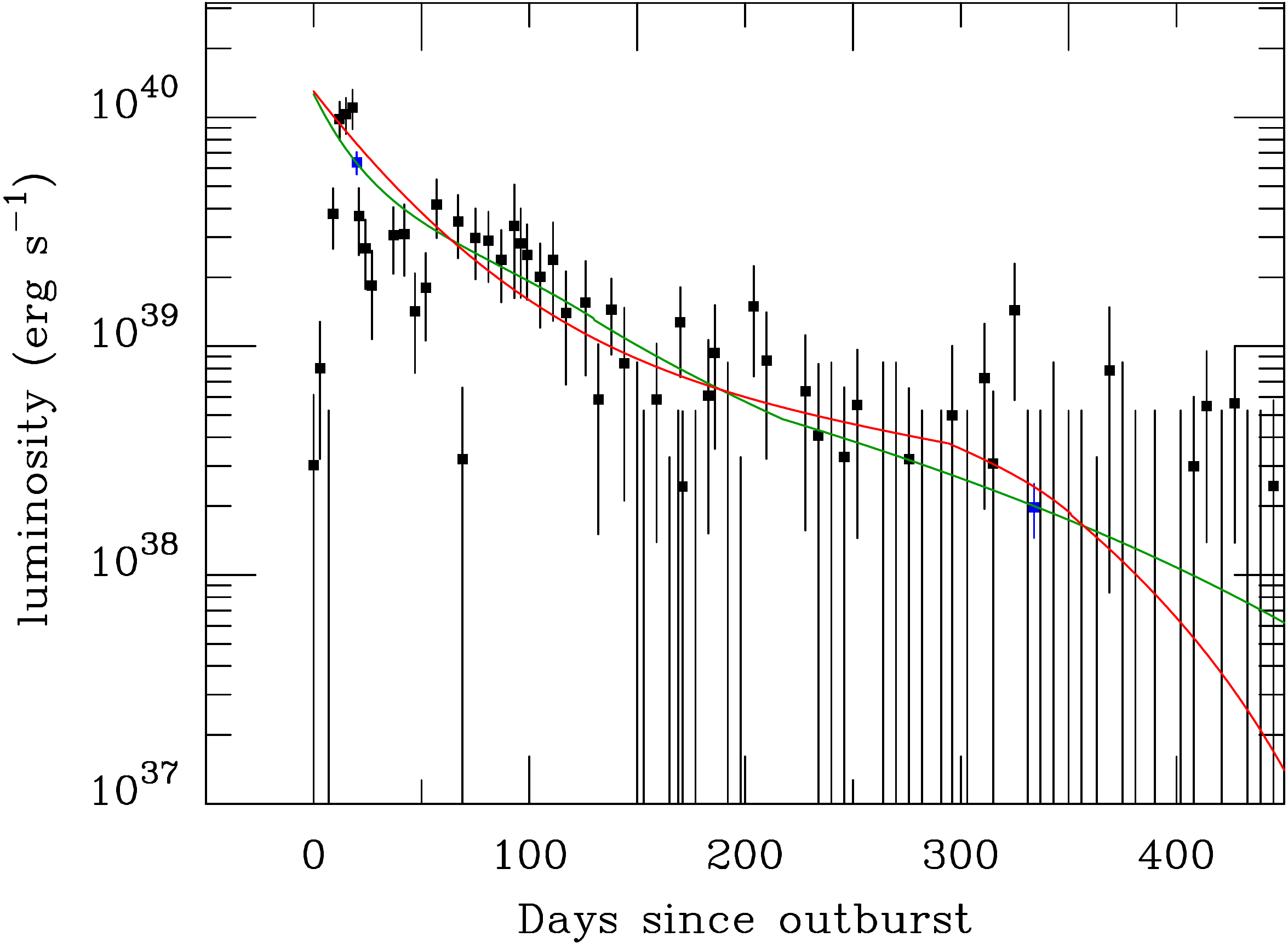}
\caption{Observed flux from M51 XT-1 as compared with model predictions for a 1.4~M$_\odot$ neutron star (red curve) and a 10~M$_\odot$ black hole (green curve). Black points correspond to {\sl SWIFT} data and blue points to {\sl Chandra} or {\sl XMM-Newton} observations \citep[courtesy M. Brigthman, see also][]{bef20}.}
\label{fig:m51_xt}
\end{figure}

\subsubsection{M51 XT-1}
Figure \ref{fig:m51_xt} shows how observational data of M51 XT-1 \citep{bef20} compare with our model when using Eqs. \ref{eq:varmdot} and \ref{eq:decay}, combined with Eq. \ref{eq:lx} with the beaming parameter $\tilde b=73$ to obtain the light curve. We have considered a 1.4 M$_\odot$ accreting neutron star, and the disc size was taken to be $4.8 \times 10^{11}$~cm. As can be seen, the agreement is reasonably good, and as acceptable as the original fit with a power law with index $-5/3$. These data can also be fitted with a 10 M$_\odot$ accreting black hole, with a disc extending to $4.9 \times 10^{11}$~cm.

In the neutron star case, the maximum accretion rate is $6 \times 10^{19}$~g\,s$^{-1}$; in the black hole case, it is $1.5 \times 10^{20}$~g\,s$^{-1}$. Hence the emission from a black-hole system would not be beamed, but in the case of a neutron-star accretor the beaming factor would be $b=0.06$, as expected for $\dot m =33$. The mass transfer rate can be very roughly estimated to be of the order of 1 -- 2\% of the maximum accretion rate, i.e. $1 - 3 \times 10^{18}$~g\,s$^{-1}$.

We checked that taking in Eq. \ref{eq:lx} $\tilde b=200$ instead of $\tilde b=73$ also provides an acceptable fit, but $\tilde b=20$ is excluded by the {\sl Chandra} and {\sl XMM-Newton} points.

\subsubsection{Other transient sources}

It is also interesting to note that the DIM predicts that, even if the mass transfer rate from the secondary is constant, sequences of outbursts with very different outburst peak luminosities and fluences can be produced. NGC 925 ULX-3 is a good example of such a situation \citep{ehb20}: this source was detected with a luminosity of $(7.8 \pm 0.8) \times 10^{39}$~erg\,s$^{-1}$ in November 2017 by {\it Chandra}; it had been detected in 2011 at a similar luminosity by {\sl SWIFT}, and also in January 2017 by {\sl XMM-Newton}, albeit at a level twenty times lower. Although the January 2017 observation does not necessarily coincides with the peak of the outburst, such a possibility should be kept in mind.

\subsubsection{Steady sources}
We suggest that some of the apparently steady ULXs could well be transient sources caught in a long-lasting outbursts. As shown earlier, outbursts can last for years. Several ULXs have been observed for a long time, and have not shown any sign of long term variability, but this option cannot be completely excluded since the DIM can easily produce outbursts lasting for 10 years (see e.g. models 22, 23, 24, 24a) or even more if the disc is larger than what we have considered here and the mass transfer rate is sufficient for the heating front to reach the outer disc edge, or if the viscosity is small (see Eqs. \ref{eq:tau} and \ref{eq:t0}). It is unclear that any of the observed steady source would fall in this category, but this option should be kept in mind when trying to account for the observed luminosity of ULXs which might not directly reflect the mass transfer rate from the secondary.

\subsubsection{HLX-1 in ESO 243-49}

HLX-1 in the galaxy ESO 243-49 \citep{fwbgr2009} is the brightest of the few hyperluminous ($L_X > 10^{41}$~erg~s$^{-1}$; one of them, NGC 5907 ULX1, with $L_X \gtrsim 10^{41}$~erg~s$^{-1}$ is a PULX) X-ray sources known; its luminosity is variable and can exceed $10^{42}$~erg~s$^{-1}$ at maximum. Its association with the galaxy ESO 243-49 at a distance of 95 Mpc is rather well established \citep{wfwsmbg2010,shp2013} so there no much doubt about the value of its luminosity \citep[see, however,][]{lkd15}. The mass of the presumed accretor is widely believed to be close to or higher than $10^4\,\rm M_{\odot}$ which would make it one of the few intermediate-mass black hole (IMBH) claimed to reside in an ULX. The once almost universal belief that all ULXs host IMBHs \citep[i.e. compact accretors with masses between $10^{2}$ and $10^{4}\, \rm M_{\odot}$;][]{cb1999} has been shattered by the discovery of an X-ray pulsar in M82 ULX-2 by \citet{betal2014}, followed by several other discoveries of such pulsing ULXs \citep[PULXs, see][for their properties and references]{kl2020}. The evidence that the compact object in HLX-1 is an IMBH is based mainly on the spectral \citep{servillat2013,godet2012,ts2016,soria2017} and radio properties \citep{webetal2012} of this source.

However, since 2008, HLX-1 has been observed to exhibit outbursts that were first appearing at about one year interval; then the recurrence became less frequent and the last outburst has been observed in April 2017 \citep[see][]{lhletal2020}. The outbursts have the typical fast-rise slow-decay (FRSD) shape of X-ray transient outbursts produced in accretion discs around stellar-mass compact accretors. Although they look like disc-instability X-ray transient events, have the same timescales (decay times of about 30 days, duration 180 days, recurrence times longer than a year or more), they cannot be their analog for an IMBH precisely because of these timescales \citep{ladetal2011}: the outbursts are much too short for a mass of $10^4\, \rm M_{\odot}$. Several models that often invoke episodic accretion from a companion star in a very eccentric orbit have been proposed to explain HLX-1 outbursts assuming that the accretor is an IMBH \citep[see e.g.][]{ladetal2011,gletal2014,lhletal2020}, but all face severe difficulties both for observational and theoretical reasons \citep[see e.g.][]{soria2017,lhletal2020,vdh2016}. 

Another model, according to which the HLX-1 outbursts would have their origin in a Compton-heated wind instability appearing at wind mass-losses much bigger than the central accretion rate \citep{shields1986} has been proposed by \citet{soria2017}. This is an interesting possibility, however, the light curves generated by such an instability do not, and cannot resemble, even remotely what is observed in HLX-1 \citep[see e.g.][]{shields1986, gp2020}.

It is not an accident since, as mentioned above, the outbursts of this hyper-luminous source have the shape naturally explained by the thermal--viscous disc instability model.
Since no other model, as yet, is able to reproduce such light curves, but the DIM applied to a disc around a $10^4\, \rm M_{\odot}$ black-hole produces wrong timescales \citep{ladetal2011}, it is legitimate to test if, at least the outburst shape, peak luminosities and characteristic times can be reproduced assuming that, as probably most of other ULXs, HLX-1 possesses a stellar-mass accretor.

In 2012, \citet{wgwetal2014} observed quasi-simultaneously in X-rays and optical the beginning of an HLX-1 outburst and found that there is a possibility that the flux in the optical V band began to rise about two days before the X-rays. Such a few-days ``X-ray delay'' is typical of X-ray transient outbursts in LMXBs \citep{russelletal2019} and is explained by the viscous-time filling of the quiescent disc's inner hole at the outburst start. Because of the short two days timescale, this cannot be the explanation if the accretor is an IMBH \citep{wgwetal2014}. In principle this X-ray delay could be explained by propagation at the sound speed of a disturbance in the disc produced by a black-hole-orbiting star, but this would just add another speculative element to models that, as mentioned above, have to face also other difficulties.

Therefore, if the ``X-ray delay" is real and produced in the disc, it would nicely fit to the other properties of the outbursts (shapes, timescales) that are typical of stellar-mass XBTs and there would be no escape from the conclusion that also here we have to do with low-mass black hole or a neutron star. Timescales are more basic than interpretation of spectra\footnote{Indeed, \citet{szzk2011} have shown for example that the X-ray spectral and timing properties of HLX-1 are equally consistent with an intermediate-mass black hole or with a foreground neutron star.}.

This has motivated \citet{lkd15} to propose a model according to which HLX-1 contains a $3\,\rm M_{\odot}$ accretor surrounded by an unstable disc whose maximum apparent luminosity $10^{42}$~erg~s$^{-1}$ corresponds to $\dot m=170$ (see Eq. \ref{eq:lx}). This solution was obtained in the framework described at the end of section \ref{sec:hot}, and we revisit it here.

Assuming that $\dot m_{\rm max} \gg 1$ and a peak luminosity of 10$^{42}$~erg\,s$^{-1}$, Eq. \ref{eq:lx} gives, in the case $\tilde b=73$: 
\begin{equation}
  \dot m_{\rm max}^2 (1 + \ln \dot m_{\rm max} ) \, M_1 = 5.6 \times 10^5,
\label{eq:dotm}
\end{equation}
which can easily be solved when $M_1$ is given. If the decay is due to the viscous decay of a disc fully in the hot state, then Eq. \ref{eq:mdot-tau_edd} must also be satisfied; the decay time varies between 18 and 62~d \citep{yzs15}; we assume here $\tau_{\rm L}  = 40$~d; this leads to
\begin{equation}
    f = 3.29 \, \dot m^{0.5}_{\rm max} M_1^{1.31} \alpha_{0.2}^{-1.34} f_{\rm irr}^{0.5} 
\end{equation}
We also requires that $f$ does not exceed 50 -- 100 for the disc to be unstable. This is equivalent to setting an upper limit on $M_1$, as it can easily be seen that $\dot m_{\rm max} M_1^{2.65}$ increases with increasing $M_1$. For a neutron star primary with $M_1 = 1.4$, $\dot m_{\rm max} = 248$, and $f=80 \alpha_{0.2}^{-1.34} f_{\rm irr}^{0.5}$. This is possible if, as in sub-Eddington transients, $\alpha_{\rm h}$ is significantly larger than 0.2. If, for example, $\alpha_{\rm h} = 0.6$ and $f_{\rm irr}=1$, one would have $f=18$. This corresponds to a peak accretion rate of $3.8 \times 10^{20}$~g\,s$^{-1}$, $\dot{M_2} \simeq 10^{19}$~g\,s$^{-1}$ and a outer disc radius of $7.9 \times 10^{11}$~cm. This is much smaller that the determination by \citet{soria2017} who found that the outer radius is about $10^{13}$~cm by fitting the optical emission with an irradiated disc model plus a red, stellar component; we do have, however, deep concerns about the accuracy of such models of the optical emission from the system in particular in the presence of strong outflows that may extend to large distances and contribute in a significant way to the optical emission. A solution with $f < 100$ with a primary black hole would imply unrealistic parameters for the viscosity and the irradiation efficiency. The difference between our conclusion and that of \citet{lkd15} is essentially due to the fact that \citet{lkd15} used the duration of the outburst instead of the decay time in the $\dot m_{\rm max} - \tau$ relation; given that $\tau_{\rm L}$ enters Eq. \ref{eq:mdot-tau_edd} to the power 3.39, the impact is quite significant. 

If, on the other hand, the observed decay is due to the propagation of a cooling front that has not (quite) reached the outer disc edge, then Eq. \ref{eq:mdot-tau_edd} does not apply, but one should use instead the relation between the the maximum accretion rate and outburst duration (Eq. \ref{eq:mdot-tp0_edd}), i.e. with $t_0^\prime = 180$~d. As for the case of the decay of a fully hot disc, the higher $M_1$, the higher is $f$. Assuming again a 1.4~M$_{\odot}$ accreting neutron star, we find that the 180~d duration can be explained with $\alpha_{0.2} = 3$ and $f_{\rm irr} = 0.5$ if $\xi = 7.7$.

The amplitude of HLX-1 outbursts is about 50, too low for a standard X-ray transient; it is not clear that the X-rays observed in quiescence are all produced by the disc, but it could also be ``fitted" by increasing $\alpha_c$ which at the same time would account for the shortish recurrence time. And there would be no problem explaining and reproducing a 2--days X-ray delay. We conclude that a stellar-mass accretor is still a viable (but admittedly not compelling) option for HLX-1 in ESO 243-49 \citep[see also][]{kl2014} which should be seriously considered since no model assuming the presence of an IMBH is able to explain the basic properties of the observed outbursts. More work is obviously needed to consolidate this option and to account for all of the observed spectral and optical properties of HLX-1, but this is a formidable task that is clearly outside the scope of this paper.

\section{Conclusions}

We extended and generalized successfully the irradiated-DIM to the case of large accretion discs and super-Eddington accretion rates.

Assuming that, during decay, the inner disc extending between the inner edge and the position of the cooling front is close to being steady with an accretion rate equal to the critical rate, we have been able to derive a relation between the peak accretion rate during an outburst and the maximum distance reached by the heating front, that closely matches the results of numerical simulations as long as the heating front does not reach the outer edge of the disc. We have also been able to solve explicitly the time evolution of the outburst decay, that consists of one or two phases: the propagation of the cooling front throughout the disc, possibly preceded by a phase during which the full disc is in the hot state. In both cases, we have been able to determine the characteristic time scales. Again, these analytic solutions are in very good agreement  with the results of numerical simulations. These results are important, because they limit the need for numerical simulations.

We have shown that, in most cases, the peak accretion rate obtained during outbursts is proportional to the mass transfer rate, with a proportionality coefficient that depends mainly on the ratio $\alpha_{\rm h}/\alpha_{\rm c}$, and is  of the order of a few tens, possibly reaching 100 when $\alpha_{\rm h}/\alpha_{\rm c} = 10$. 

We found that sub-Eddington outbursts of systems with large orbital periods, such as V404 Cyg are well accounted for by the DIM. We have shown that, provided that the fraction of the X-ray flux that irradiates the accretion disc increases when the X-ray luminosity exceeds the Eddington luminosity and hence the accretion efficiency decreases, super-Eddington accretion rates can be obtained in large accretion discs during outbursts. We have shown that ULX transients (ULXTs), such as M51 XT-1 are well-described by the DIM with parameters consistent with the general properties of such sources. Some of these outbursts might last several years, and could thus account for some other types of variability of observed ULXs. Outbursts of HLX-1 in ESO 243-49 can be described by our model, on the condition that the accretor is a neutron star.

In our models, a number of transient ULXs are assumed to be beamed. This certainly must be the case for ULXs containing neutron stars, since luminosities exceeding up to one hundred times the Eddington luminosity cannot be attained if $L_{\rm x}$ increases as $\ln \dot m$. Other options have been suggested, such as for example the presence of a magnetar type magnetic field, but they all are not viable \citep[see][for a discussion of these]{kl2020}. Beaming implies that there exists a large number of sources that would be seen off axis, and therefore would not appear as ULXs, although they would be surrounded by a nebula, such as SS~433 that is most probably an ULX seen from the side.

We also note that not all ULXs need to be beamed since according to Eq. \ref{eq:lx} this should the case only for $\dot m > \sqrt{\tilde b} \gtrsim 10$. Indeed, using simulations of stellar populations, \citet{wgl19} have shown that observed ULXs harbouring black hole accretors are typically emitting isotropically, whereas systems containing neutron stars would be beamed. This is in line with observational arguments showing that, at least in some systems, the luminosity averaged over solid angle is not very different from the apparent luminosity. For example, a comparison of the luminosities from the soft X-rays and from the helium line emitted by the surrounding nebula suggests that the beaming effect is at best moderate in a few sources \citep{pm02,mhc11,ryg11}; NCG 300 ULX-1, which contains a neutron star, is in a similar situation \citep{bld18}, but since its apparent luminosity is $4.4 \times 10^{39}\rm erg/s$ it corresponds according $b\approx 0.2$ \citep{bld18,kl2019}. However, bright PULXs with $ > 10^{40}\rm erg/s$ must be strongly beamed since no physically motivated alternative exists.

We finally note that in our model a potentially crucial ingredient, namely winds, is included only implicitly and in a very crude way. Winds are required here to avoid the decrease in the illumination factor $f_{\rm irr}$ when the accretion rate exceeds the Eddington value, but they also modify significantly the outburst duration and could have a major impact on angular momentum transport if the wind is magnetized. Unfortunately, although they have an important effect on the whole outburst cycle, the physics of outflows from accretion discs are still not fully understood \citep{ddt19,tdmdc2020}.

Thus the DIM applied to ULXTs suffers from the same weaknesses as the ``standard'' DIM used to describe LMXB transient sources (some are even enhanced) but, as the latter model, has the advantage to be based on sound physical assumptions and could, in principle, be easily tested by observations, although, in practice the large distances to ULXs are an obvious drawback.

\begin{acknowledgements}
We thank Murray Brightman for having provided us with the observational data points of M51 XT-1. We are grateful to Guillaume Dubus, Andrew King, and Matt Middleton for helpful comments and suggestions. JPL was supported by a grant from the French Space Agency CNES. 
\end{acknowledgements}

\bibliographystyle{aa}
\bibliography{large_discs}

\end{document}